\documentclass[11pt,singlecolumn,nofootinbib,notitlepage,superscriptaddress]{revtex4-1}

\usepackage{amssymb,amsmath,amsfonts}
\usepackage[usenames]{color}
\usepackage[pdftex]{graphicx}
\usepackage[english]{babel}
\usepackage{multirow}
\usepackage{txfonts}

\usepackage[percent]{overpic}
\usepackage{overpic}
\usepackage{mathrsfs}

\usepackage{array, enumerate}
\usepackage[breaklinks,colorlinks,citecolor=blue]{hyperref}
\usepackage{enumitem}
\usepackage{bm}
\usepackage{braket}

\begin{document}
\title{A Note on QPE Timing: False Alarms in O-C}

\author{Cong Zhou}
\email{dysania@mail.ustc.edu.cn}
\affiliation{Department of Astronomy, University of Science and Technology of China, Hefei 230026, People’s Republic of China}

\author{Zirui Zhang}
\email{zirui379@whu.edu.cn}
\affiliation{School of Physics and Technology, Wuhan University, Wuhan 430072, People’s Republic of China}

\author{Zhen Pan}
\email{zhpan@sjtu.edu.cn}
\affiliation{Tsung-Dao Lee Institute, Shanghai Jiao-Tong University, Shanghai, 520 Shengrong Road, 201210, People’s Republic of China}

\author{Wen Zhao}
\email{wzhao7@ustc.edu.cn}
\affiliation{Department of Astronomy, University of Science and Technology of China, Hefei 230026, People’s Republic of China}

\date{\today}

\begin{abstract}
O-C timing analysis is a useful diagnostic tool for quasi-periodic eruptions (QPEs), but their interpretation depends sensitively on the integer cycle number assigned to each eruption. In this note, we show that even a small mismatch in the cycle number, $N_{\rm cyc}$, can produce large false signals in O-C diagrams, and \emph{ a universal feature of these false signals 
is a large in-phase sinusoidal modulation between even and odd eruptions.}  Therefore,   uncertainties in $N_{\rm cyc}$  must be inferred or marginalized over before physical interpretations are attached to O-C.  %Using controlled mock data, we compare the signatures of apsidal precession, disk precession, and incorrect cycle assignments.  
We then apply both O-C and EMRI+disk to GSN 069 and eRO-QPE2. 
For GSN 069, the timing data favor an anti-phase  modulation in even and odd eruptions, consistent with apsidal precession in a low-eccetricity EMRI crossing an equatorial disk.  For eRO-QPE2, the data are well described by a near-circular EMRI and a precessing disk. %, while the cycle assignment $N_{\rm FinalEpochStart}=323$ together with a tiny negative period derivative is not preferred by the timing data.  
%The main methodological conclusion is that cycle-number uncertainties must be inferred or marginalized over before physical interpretations are attached to O--C modulations.
\end{abstract}

\maketitle

\section*{Overview and reading guide}

The main point of this note is simple: an O-C analysis is useful only after the integer cycle number of each eruption, $N_{\rm cyc}$, has been handled carefully.  Here ``O-C'' means
\begin{equation}
    {\rm O-C} = t_{\rm obs} - t_{\rm calc},
\end{equation}
where $t_{\rm obs}$ is the observed eruption time and $t_{\rm calc}$ is the time predicted by a toy timing model.  The calculated time $t_{\rm calc}(N_{\rm cyc}, \mathbf{\Theta})$ depends not only on model parameters $\mathbf{\Theta}$, but also on the assumed cycle number $N_{\rm cyc}$.  As a result,  a small error in $N_{\rm cyc}$ leads to false signals in O-C.

The most important diagnostic for O-C in this note is the phase relation between the even and odd eruptions:
\begin{itemize}[leftmargin=*]
    \item An eccentric EMRI crossing an equatorial disk usually produces an {\bf anti-phase} modulation between even and odd eruptions.  This is equivalent to the alternating long-short pattern: alternating  $T_{\rm long}$ and $T_{\rm short}$, $T_{\rm long}$ and $T_{\rm short}$ vary with time, while $T_{\rm long}(t)+T_{\rm short}(t)$ remains approximately constant.
    \item A precessing disk can produce an {\bf in-phase} modulation between even and odd eruptions.
    \item An incorrect choice of $N_{\rm cyc}$ also produces a large {\bf in-phase} modulation.  This is the main false-alarm mechanism discussed below.
\end{itemize}

Thus, an in-phase modulation should not immediately be interpreted as  disk precession (or a second SMBH).  One must first check whether the same feature is due to a small error in the cycle number assignment.

% The O-C (“observed” timing -“calculated” timing based on a toy model with a constant period or a constant period derivative) analysis is very sensitive to the cycle number $N_{\rm cyc}$ of each eruption. A small error in $N_{\rm cyc}$ leads to large false signals, and a universal signature of these false signals is a large in-phase sinusoidal modulation in even and odd data.   Therefore, uncertainties in $N_{\rm cyc}$ must be properly quantified for making solid claims based on O-C.
% As examples, we  show the O-C analyses on  mock data, GSN 069 and eRO-QPE2 real data as follows. 

\section*{Notation}

For clarity, we summarize the main symbols used throughout this note.  In particular, $\dot T$ and $\dot T_{\rm obt}$ should not be identified blindly: $\dot T$ denotes a phenomenological parameter in  O-C toy models, while $\dot T_{\rm obt}$ denotes the orbital-period derivative in the EMRI + disk model.

\begin{table}[h]
\caption{Summary of notation used in this note.}
\label{tab:notation}
\small
\renewcommand{\arraystretch}{1.15}
\newcommand{\mcell}[1]{\parbox[t]{0.68\columnwidth}{\raggedright #1}}

\begin{tabular}{ll}
\hline\hline
Symbol & Meaning \\
\hline
$N_{\rm cyc}$ & \mcell{Integer cycle number assigned to an eruption. This number connects eruptions separated by observational gaps.} \\
$N_{\rm FinalEpochStart}$ & \mcell{Starting cycle number of the final observing epoch. For eRO-QPE2, this refers to the last XMM epoch considered in the timing analysis.} \\
$P_1, A_1$ & \mcell{Period and amplitude of the short period phenomenological sinusoidal modulation in an O-C toy model.} \\
$P_2, A_2$ & \mcell{Period and amplitude of the long period phenomenological sinusoidal modulations in an O-C toy model. } \\
$\dot T$ & \mcell{Phenomenological period derivative in an O-C toy model, such as a quadratic baseline fit.} \\
$\dot T_{\rm obt}$ & \mcell{Orbital-period derivative in the EMRI + disk model. It is model-dependent and is not identical to the toy-model parameter $\dot T$.} \\
$T_{\rm obt}$ & \mcell{Orbital period of the EMRI system used in the physical timing model.} \\
$e$ & \mcell{Orbital eccentricity of the EMRI.} \\
$T_{\rm aps}$ & \mcell{Apsidal-precession period of the EMRI orbit. In an eccentric EMRI crossing an equatorial disk, it usually produces an anti-phase modulation between even and odd eruptions.} \\
$\tau_{\rm p}$ & \mcell{Disk-precession period. A precessing disk can produce an in-phase modulation between even and odd eruptions.} \\
$T_{\rm long}, T_{\rm short}$ & \mcell{Alternating long and short recurrence times between successive eruptions. Their time dependence is another way of describing the anti-phase even-odd O-C modulation.} \\
$\phi_{\rm even}, \phi_{\rm odd}$ & \mcell{Phases of the fitted modulation in the even and odd eruption sequences. Anti-phase means $\phi_{\rm even}-\phi_{\rm odd}\simeq\pi$, while in-phase means $\phi_{\rm even}-\phi_{\rm odd}\simeq0$.} \\
$\sigma_{\rm sys}$ & \mcell{Additional systematic timing uncertainty used in the EMRI + disk fit.} \\
\hline\hline
\end{tabular}
\end{table}

\newpage

\section*{I. Circular EMRI + equatorial disk}
{\bf Injected model:} $T_{\rm obt}=64.72\ {\rm ks}$, $\dot T_{\rm obt}=0$, and $e=0$.

\emph{Purpose of this test.}
This is the cleanest possible mock example.  The orbit is circular, the disk is fixed, and the orbital period does not evolve.  Therefore, the true timing data  contain no apsidal-precession signal, no disk-precession signal, and no period derivative.

% {\bf I. Circular EMRI + Equatorial disk (with $T_{\rm obt} =64.72\ {\rm k s}, \dot T_{\rm obt} =0, e=0$)}

\begin{figure}[h]
    \centering
    \includegraphics[width=0.45\linewidth]{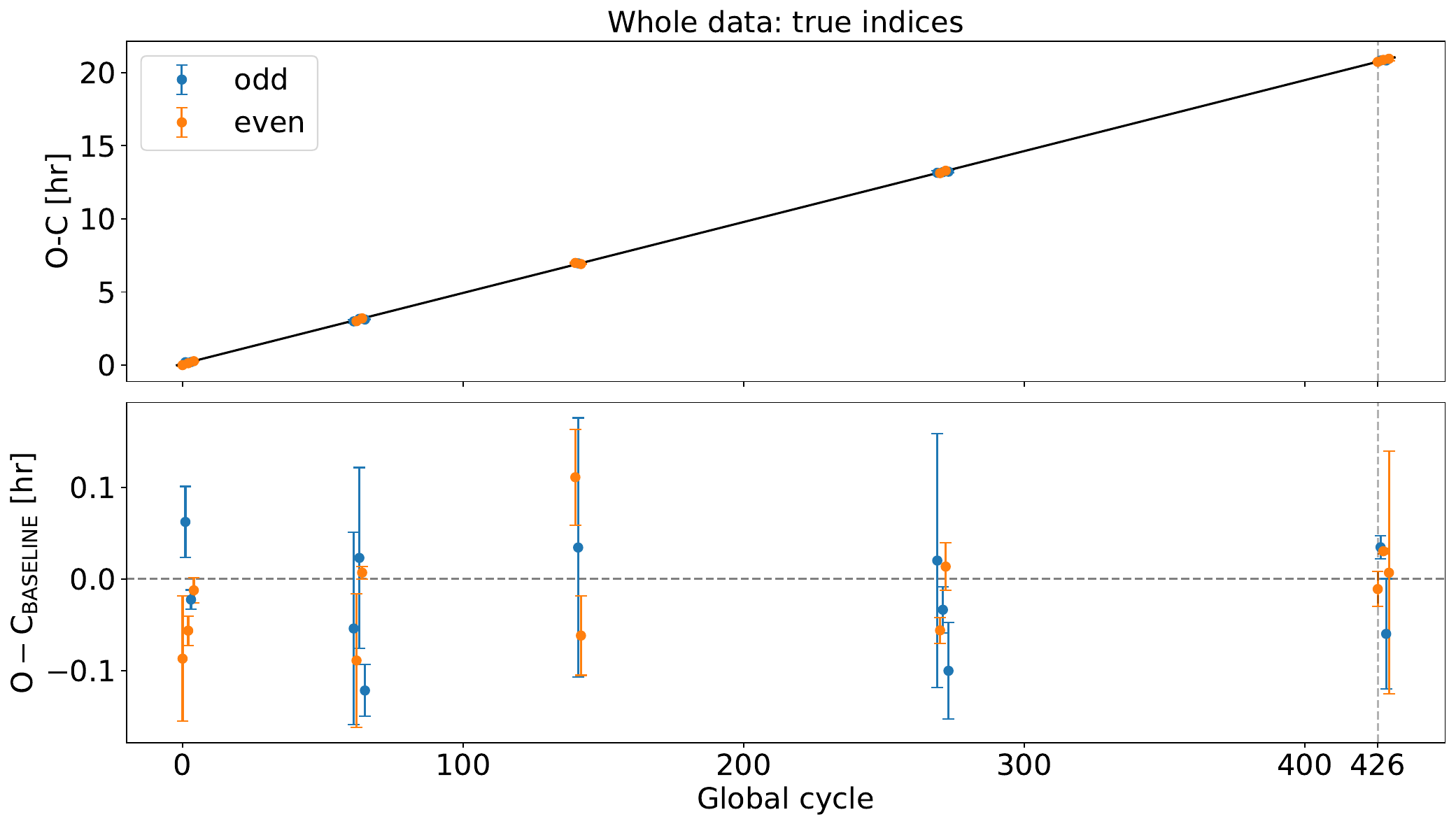}
    \includegraphics[width=0.45\linewidth]{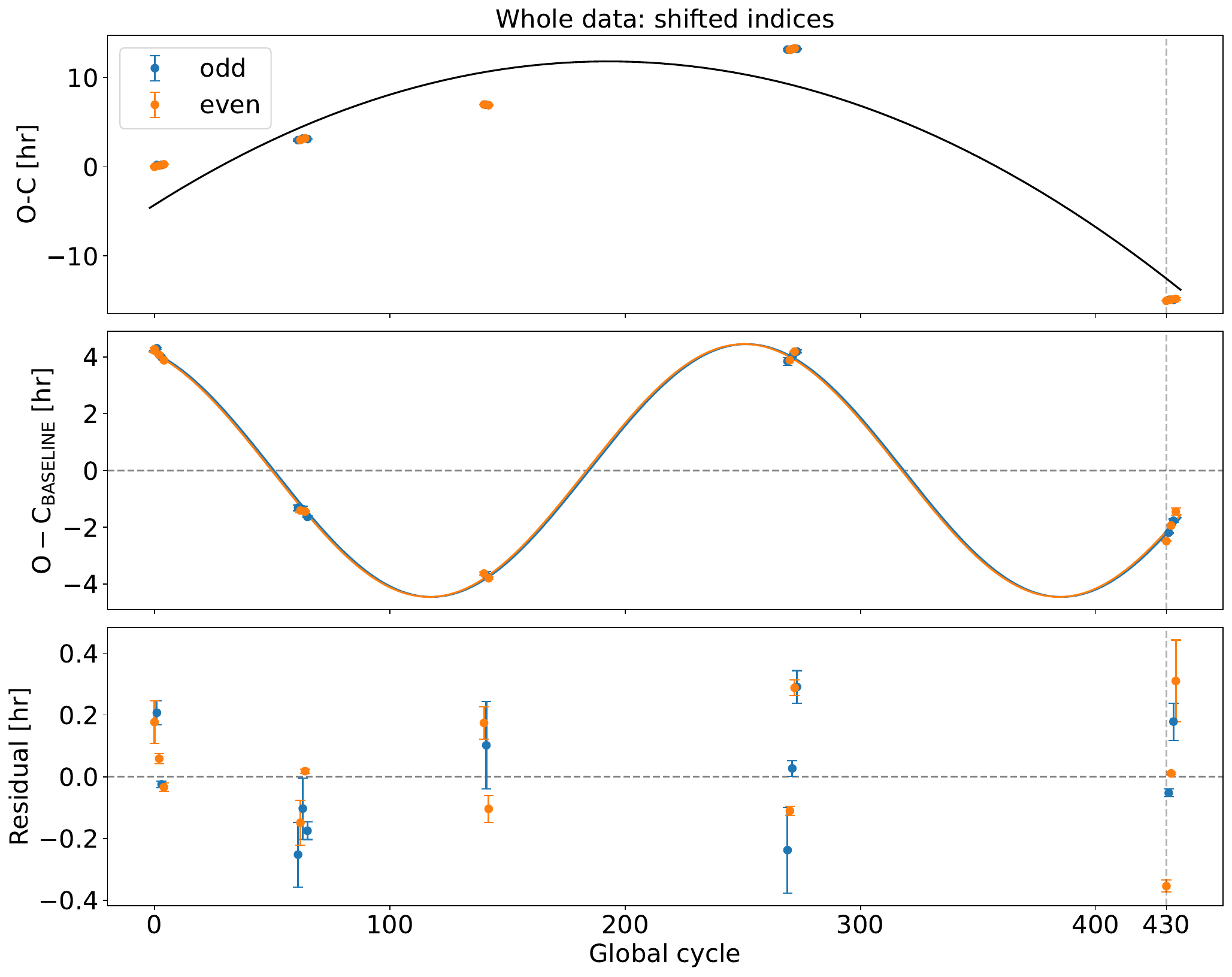}
    \caption{Circular mock data. Left: O--C fit with the correct cycle assignment. Right: O--C fit after shifting the cycle number of the last epoch.  The right panel shows how a small error in $N_{\rm cyc}$ creates artificial curvature and a large sinusoidal modulation.}
    \label{fig:mock_fi}
\end{figure}

\emph{Results.}
With the correct $N_{\rm cyc}$, a linear timing model is sufficient.  After subtracting this baseline, the residuals are consistent with noises.  In contrast, after shifting $N_{\rm FinalEpochStart}$, the same data require a quadratic trend plus a large sinusoidal modulation.

~\\

\emph{Lesson.}
A incorrect cycle assignment leads to two false signals:
\begin{enumerate}
    \item an artificial non-zero $\dot T$.
    \item an artificial large {\bf in-phase} sinusoidal modulation in even and odd data.
\end{enumerate}
% This example contains no real physical modulation, so the right panel is entirely a false alarm caused by the incorrect $N_{\rm cyc}$.

% Conclusions:
% \begin{enumerate} 
%     \item Correct O-C recovers the injected linear relation.  
%     \item Incorrect O-C with a small error in cycle number $N_{\rm cyc}$ leads to {\bf two false signals}: (1) non-zero $\dot T_{\rm obt}$ and (2) a large {\bf in-phase sinusoidal modulation} in {\bf even and odd data}.
% \end{enumerate} 
% A large   in-phase sinusoidal modulation in even and odd data is  a universal feature for incorrect O-C as shown later.

% A question is then how to tell or quantify which O-C is favored in the case of real data, where no inject value is available ? 

% \begin{enumerate}
%     \item Two-parameter fit $C_{\rm linear}$ in the left panel reveals much lower $\chi^2$ than the five-parameter $C_{\rm quadratic+ mod}$ in the right panel: $\chi^2_{\rm linear}=78.34 \ll \chi^2_{\rm quadratic+mod}=735.90$ exclude $N_{\rm FinalEpochStart}=430$. 
%     % reduced $\chi^2_{\rm linear}=3.73$ and reduced $\chi^2_{\rm quadratic+mod}=45.99$ exclude $N_{\rm FinalEpochStart}=430$. 
%     \item Log Bayes factor ($\log \mathcal{B}^{\rm linear}_{\rm quadratic+mod}=157.91 \gg 1$) excludes $N_{\rm FinalEpochStart}=430$.
% \end{enumerate}

\newpage

\section*{II. Eccentric EMRI + equatorial disk, without period decay}
\textbf{Injected model:} $T_{\rm obt}=64.72\ {\rm ks}$, $\dot T_{\rm obt}=0$, and $e=0.06$.

\emph{Purpose of this test.}
%This example adds eccentricity.  
Apsidal precession of an eccentric EMRI results in an anti-phase modulation between even and odd eruptions.
%Eccentricity makes the disk-crossing times sensitive to apsidal precession, so the 
Correct O-C should reveal this anti-phase modulation. % between even and odd eruptions.

% {\bf II. Eccentric EMRI + Equatorial disk (with $T_{\rm obt} =64.72\ {\rm k s}, \dot T_{\rm obt} =0, e=0.06$)}

\begin{figure}[h]
    \centering
    \includegraphics[width=0.45\linewidth]{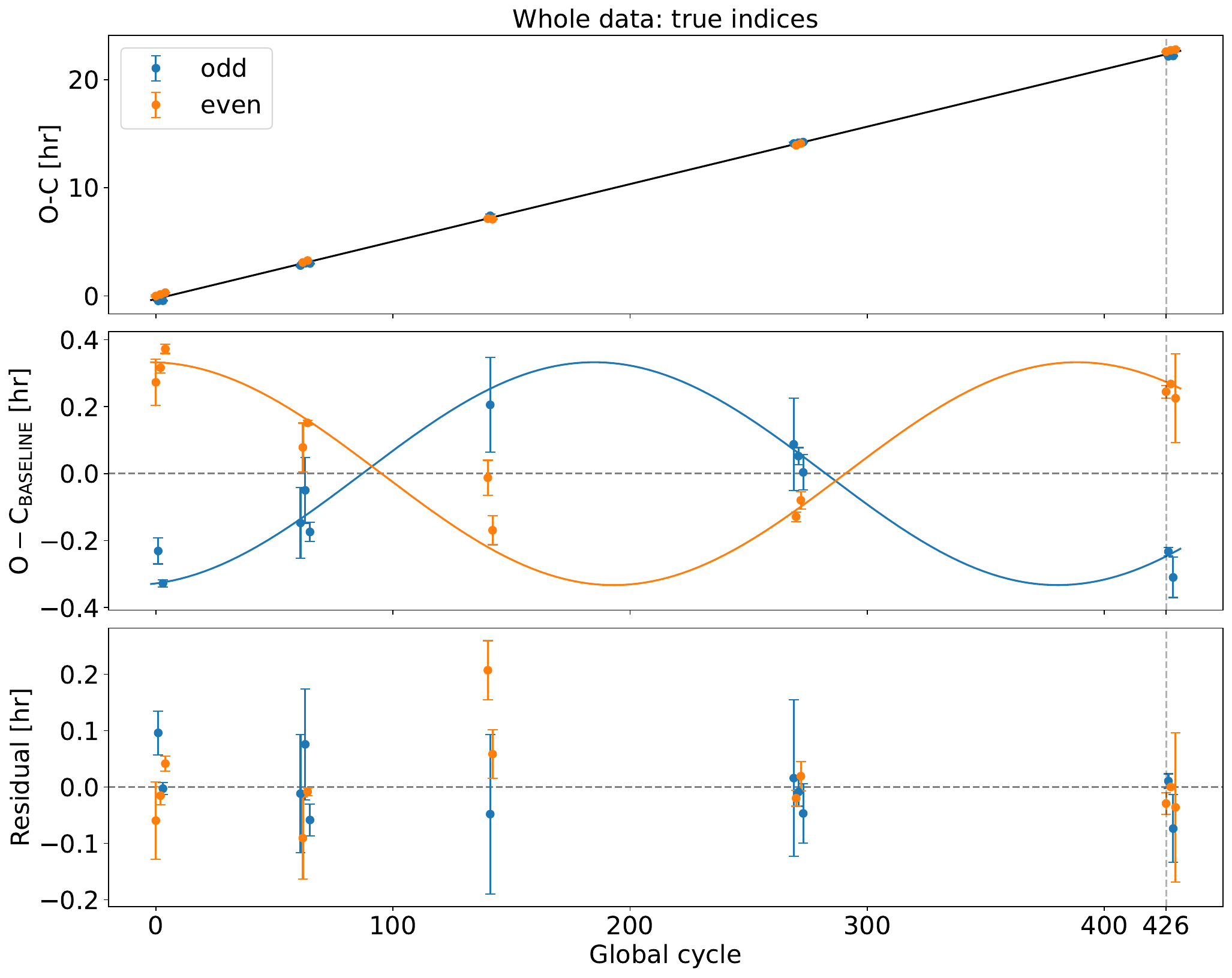}
    \includegraphics[width=0.45\linewidth]{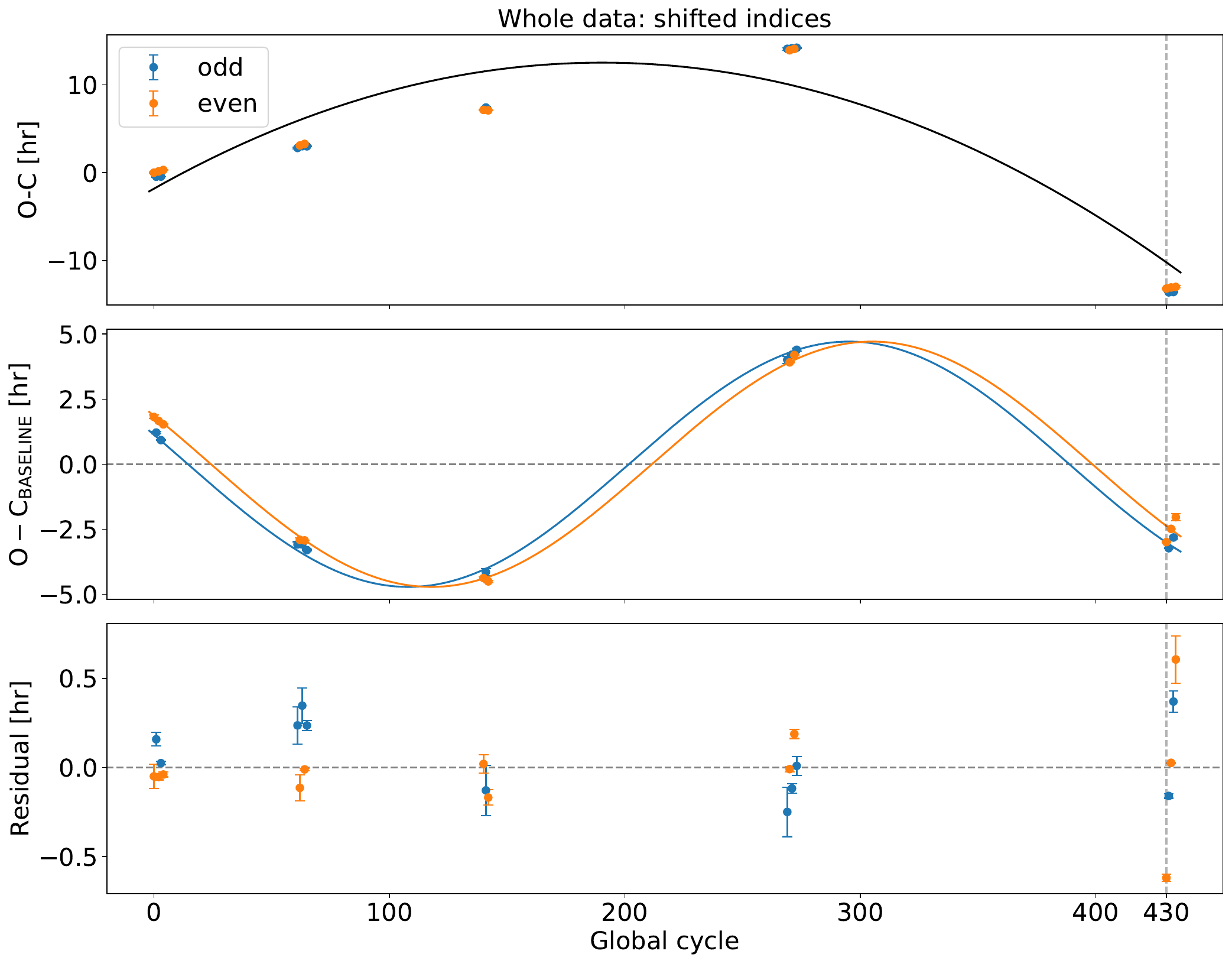}
    \caption{Same as Fig.~\ref{fig:mock_fi}, but for an eccentric EMRI.  The correct O--C fit recovers the anti-phase modulation caused by apsidal precession.  The shifted cycle assignment instead produces a much larger in-phase modulation.}
    \label{fig:mock_geo}
\end{figure}

\emph{Results.}
With correct cycle assignment, O-C recovers the anti-phase modulation in even and odd  data. % When one branch moves upward, the other moves downward.  
This is the expected signature of EMRI apsidal precession. % in the EMRI + equatorial-disk picture.
With the shifted $N_{\rm FinalEpochStart}$, O-C reveals a large in-phase modulation in the even and odd data.  

~\\

\emph{Lesson.}
The well known {\bf alternating long-short pattern} $\Big($alternating $T_{\rm long}, T_{\rm short}$ $\&$ $T_{\rm long, short}(t)$ are time-dependent  $\&$ $T_{\rm long}(t)+T_{\rm short}(t)\approx$ const$\Big)$ is an equivalent expression to the {\bf anti-phase} modulation in even and odd data.  %They are two equivalent ways of describing the same timing behavior.  Therefore, if 
An O-C fit that fails to recover the anti-phase modulation is incorrect.

% \begin{enumerate} Conclusions:
%     \item Correct O-C recovers the injected parameters and the {\bf anti-phase sinusoidal modulation} driven by the EMRI apsidal precession in even and odd data. 
%     \item Incorrect O-C with a small error in cycle number $N_{\rm cyc}$ leads to {\bf two false signals}: (1) non-zero $\dot T_{\rm obt}$ and (2) a large {\bf in-phase sinusoidal modulation} in even and odd data. The false in-phase modulation is much stronger than the true anti-phase signal.
% \end{enumerate} 

% In fact, {\bf alternating long-short pattern} $\Big($alternating $T_{\rm long}, T_{\rm short}$ $\&$ $T_{\rm long, short}(t)$ are time-dependent  $\&$ $T_{\rm long}(t)+T_{\rm short}(t)\approx$ const$\Big)$ is an equivalent expression to the {\bf anti-phase} modulation in even and odd data.  The O-C in the right panel is  incorrect simply because it does not reveal the anti-phase modulation.

% {\bf add a plot of $t_{n+1}-t_n}$ v.s. $t$, even and odd in different colors}

% \begin{figure}[h]
%     \centering
%     \includegraphics[width=0.7\linewidth]{Papp_geo.pdf}
%     \caption{The apparent period.}
%     \label{fig:Papp_geo}
% \end{figure}

\newpage 

% {\bf III. Eccentric EMRI + Equatorial disk (with $T_{\rm obt} =64.72\ {\rm k s}, \dot T_{\rm obt} =-6.5\times10^{-5}, e=0.06$)}

\section*{III. Eccentric EMRI + equatorial disk, with period decay}
\textbf{Injected model:} $T_{\rm obt}=64.72\ {\rm ks}$, $\dot T_{\rm obt}=-6.5\times10^{-5}$, and $e=0.06$.

\emph{Purpose of this test.}
This example is closer to a realistic timing analysis because the injected orbit has both eccentricity and a non-zero period derivative.  The purpose is to show that a visually good O-C fit does not by itself prove that the cycle assignment is correct.

\begin{figure}[h]
    \centering
    \includegraphics[width=0.45\linewidth]{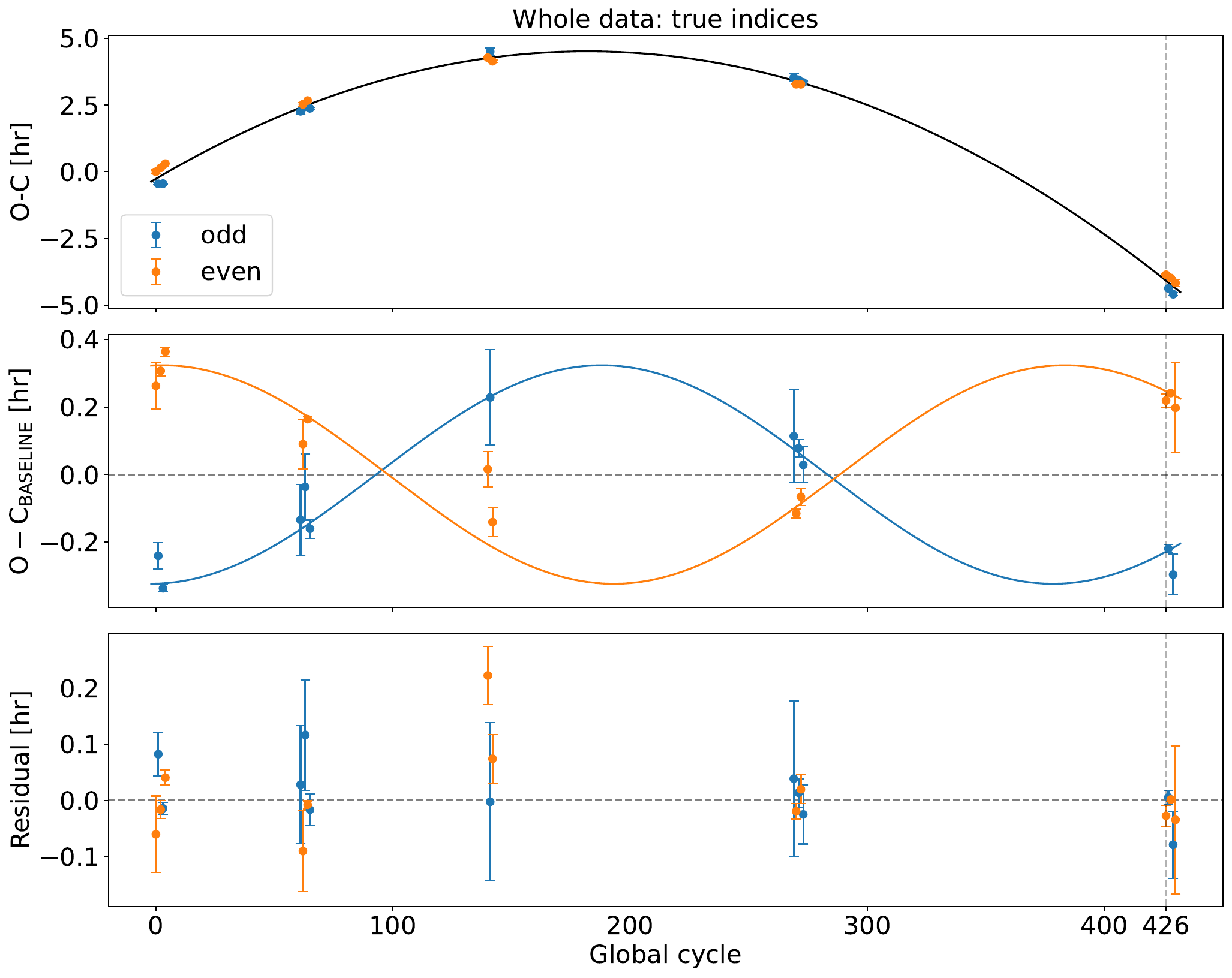}
    \includegraphics[width=0.45\linewidth]{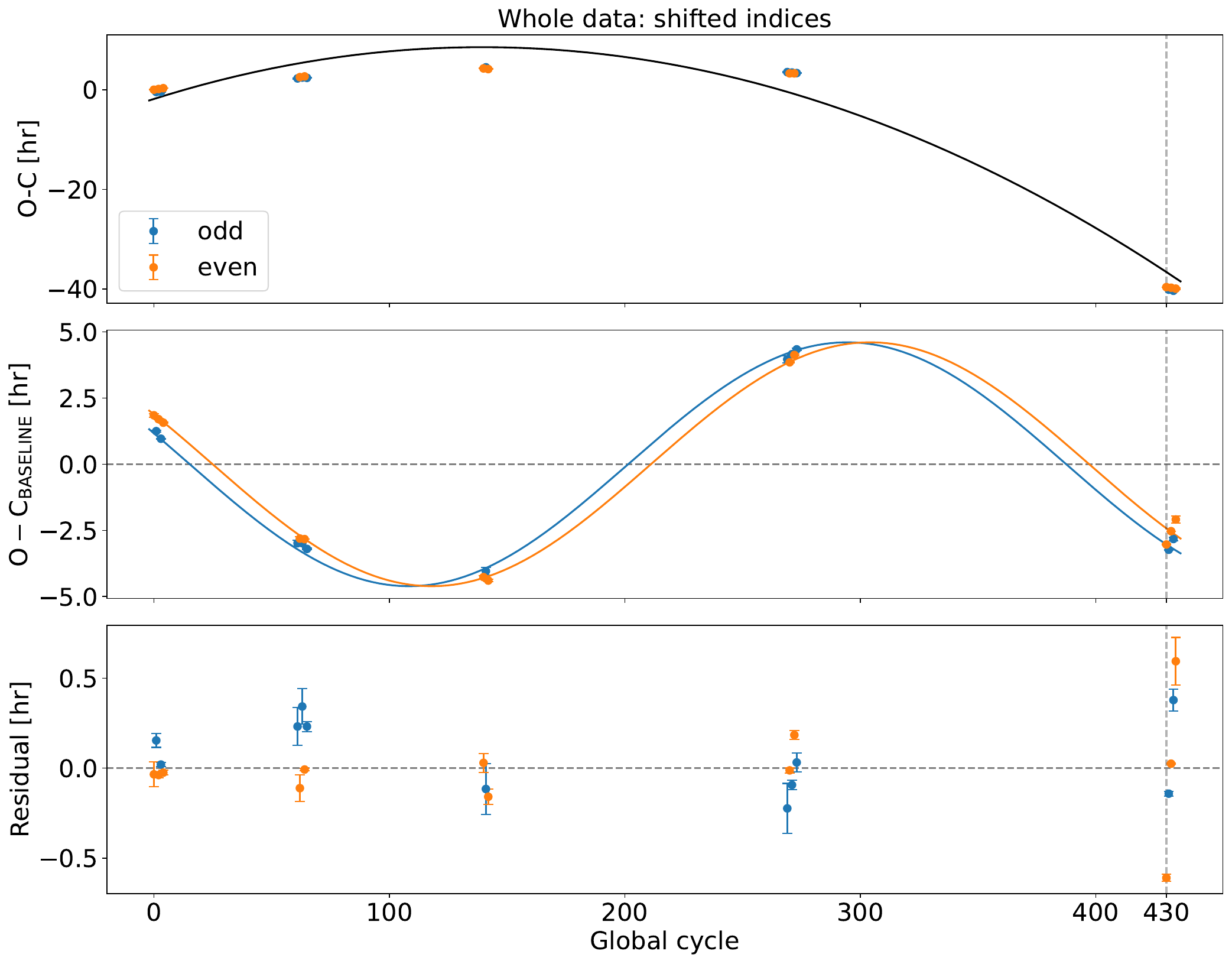}
    \caption{Same as Fig.~\ref{fig:mock_fi}, but the mock data now include a true constant $\dot T_{\rm obt}$.  Both the correct and incorrect cycle assignments can be fitted by a quadratic baseline plus a sinusoidal modulation.}
    \label{fig:mock_Tdot}
\end{figure}

\emph{Results.}
The correct O-C recovers an anti-phase modulation in even and odd data, while the incorrect one reveals an in-phase modulation.
% Because the true signal already contains a quadratic timing trend, both the correct and incorrect O-C diagrams look acceptable after fitting a quadratic baseline plus a sinusoidal modulation.  
%The difference is not simply whether the residuals look small.  The key question is whether the fitted phase relation and the cycle assignment are physically and statistically consistent.

~\\

\emph{Lesson.}
A good-looking O-C residual plot is not enough. The uncertainty in $N_{\rm cyc}$ must be quantified before interpreting O -- C results. 
%When $\dot{T}_{\rm obt}$ and sinusoidal modulations are both allowed, the model can absorb cycle-indexing mistakes.  This makes it essential to quantify the uncertainty in $N_{\rm cyc}$ rather than fixing it by hand.

\newpage 

% {\bf IV. Near-circular orbit + Precessing disk (with $T_{\rm obt} =64.72\ {\rm k s}, \dot T_{\rm obt} =-6.5\times10^{-5}, e=10^{-5}$ and a disk precession period $\tau_{\rm p}=50\ {\rm d}$)}

\section*{IV. Near-circular orbit + precessing disk}
\textbf{Injected model:} $T_{\rm obt}=64.72\ {\rm ks}$, $\dot T_{\rm obt}=-6.5\times10^{-5}$, $e=10^{-5}$, and disk precession period $\tau_{\rm p}=50\ {\rm d}$.

\emph{Purpose of this test.}
This example shows the case where an in-phase modulation is real.  The EMRI orbit is nearly circular, so the apsidal precession imprint is weak.  The dominant super-orbital timing signal comes from disk precession, which naturally leads to an in-phase modulation. % even and odd disk crossings in phase.

\begin{figure}[h]
    \centering
    \includegraphics[width=0.45\linewidth]{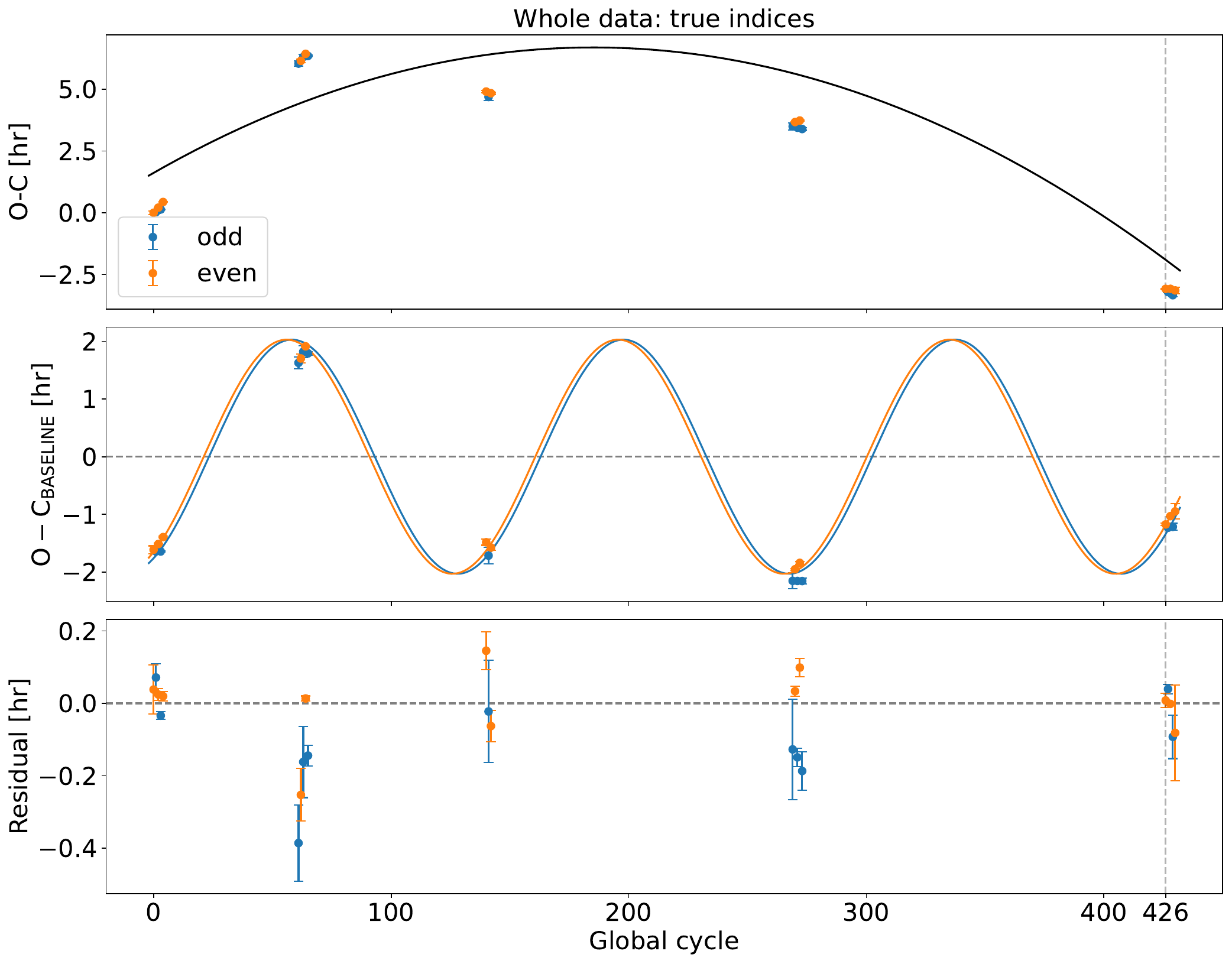}
    \includegraphics[width=0.45\linewidth]{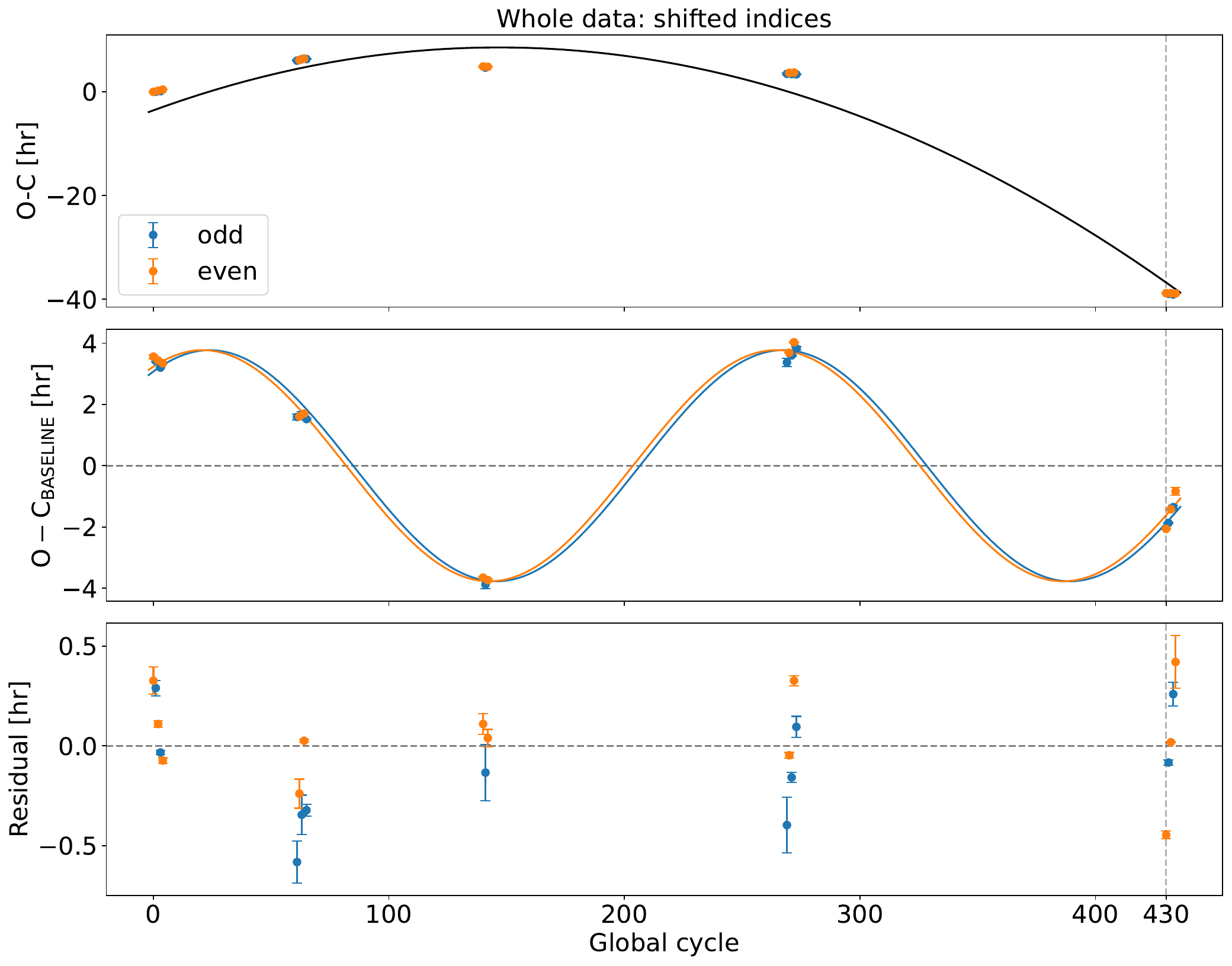}
    \caption{Near-circular EMRI with a precessing disk.  The correct O--C fit recovers the injected disk-precession signal.  The shifted cycle assignment still produces an in-phase modulation, but with biased amplitude and period.}
    \label{fig:precessing_disk}
\end{figure}

\emph{Results.}
With the correct cycle assignment in O -- C, the recovered best-fit disk-precession period is $\tau_{\rm p}= 52\ {\rm d}$, close to the injected value of $50\ {\rm d}$.  With the shifted $N_{\rm FinalEpochStart}$, the fitted signal remains in phase, but the amplitude is twice higher than the injection value, about $4\ {\rm h}$, and the period is biased to $\tau_{\rm p}=91\ {\rm d}$.

~\\

\emph{Lesson.}
An in-phase modulation can be physical, e.g., in the case of a near-circular EMRI with a precessing disk.  However, the same pattern can also be produced by a wrong $N_{\rm cyc}$. 
The incorrect O-C leads to incorrect inference of the period evolution $\dot T$, and the amplitude and period of the sinusoidal modulation.  
%Therefore, the amplitude and period of an in-phase signal are reliable only if the cycle assignment has been treated self-consistently. 

% Near-zero eccentricity (weak alternating long-short pattern, like eRO-QPE2) leaves few imprints of apsidal precession in the O–C data. In such cases, the super-orbital modulation can instead be dominated by a precessing disk. The resulting disk-precession signal is, of course, {\bf in-phase} and has a much larger amplitude than that induced by apsidal precession.

% \begin{enumerate} Conclusions:
%     \item Correct O-C recovers injected parameters and the {\bf in-phase sinusoidal modulation} driven by the disk precession in {\bf even and odd data}. In this particular example, the best-fit is $\tau_p=52\ {\rm d}$, pretty close to the injection. In more general case (e.g. more sparse data), it should be determined up to an integer multiple.
%     \item Incorrect O-C again reveals a in-phase sinusoidal modulation in even and odd data, but the amplitude obtained ($\approx 4$ h) is  twice higher than the injection value and the period is also biased. In this particular example, the best-fit is $\tau_{\rm p}=91\ {\rm d}$. 
% \end{enumerate}

\newpage

% {\bf V. GSN 069 real data}
\section*{V. GSN 069 real data}

\emph{Purpose of this section.}
GSN 069 provides a real-data example where the anti-phase modulation provides direct evidence for EMRI apsidal precession.
%and in-phase modulations is important.  %The question is whether the data favor the physically expected anti-phase apsidal-precession signal or a large in-phase modulation.

\begin{figure}[h]
    \centering
    \includegraphics[width=0.45\linewidth]{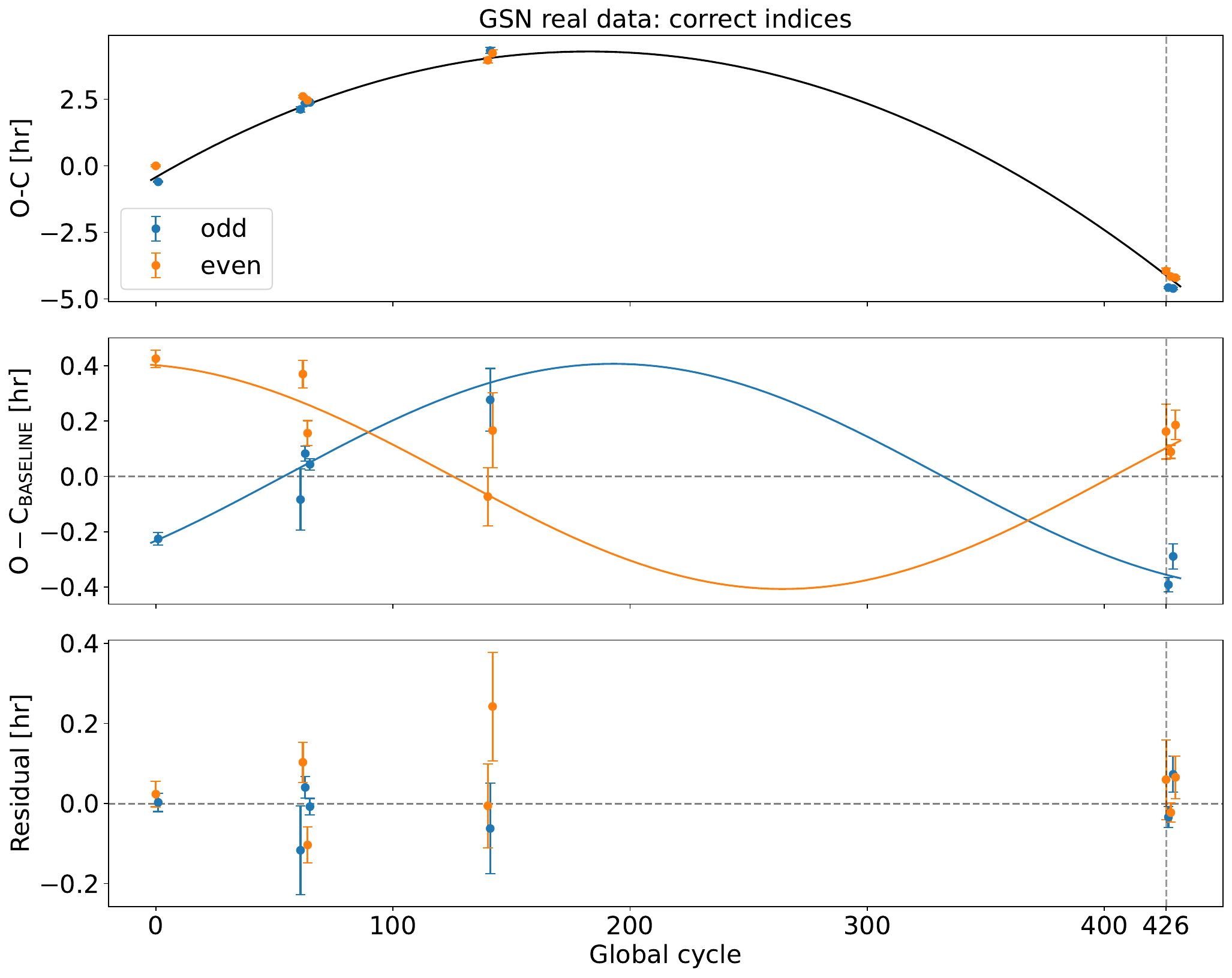} \\ 
    \includegraphics[width=0.45\linewidth]{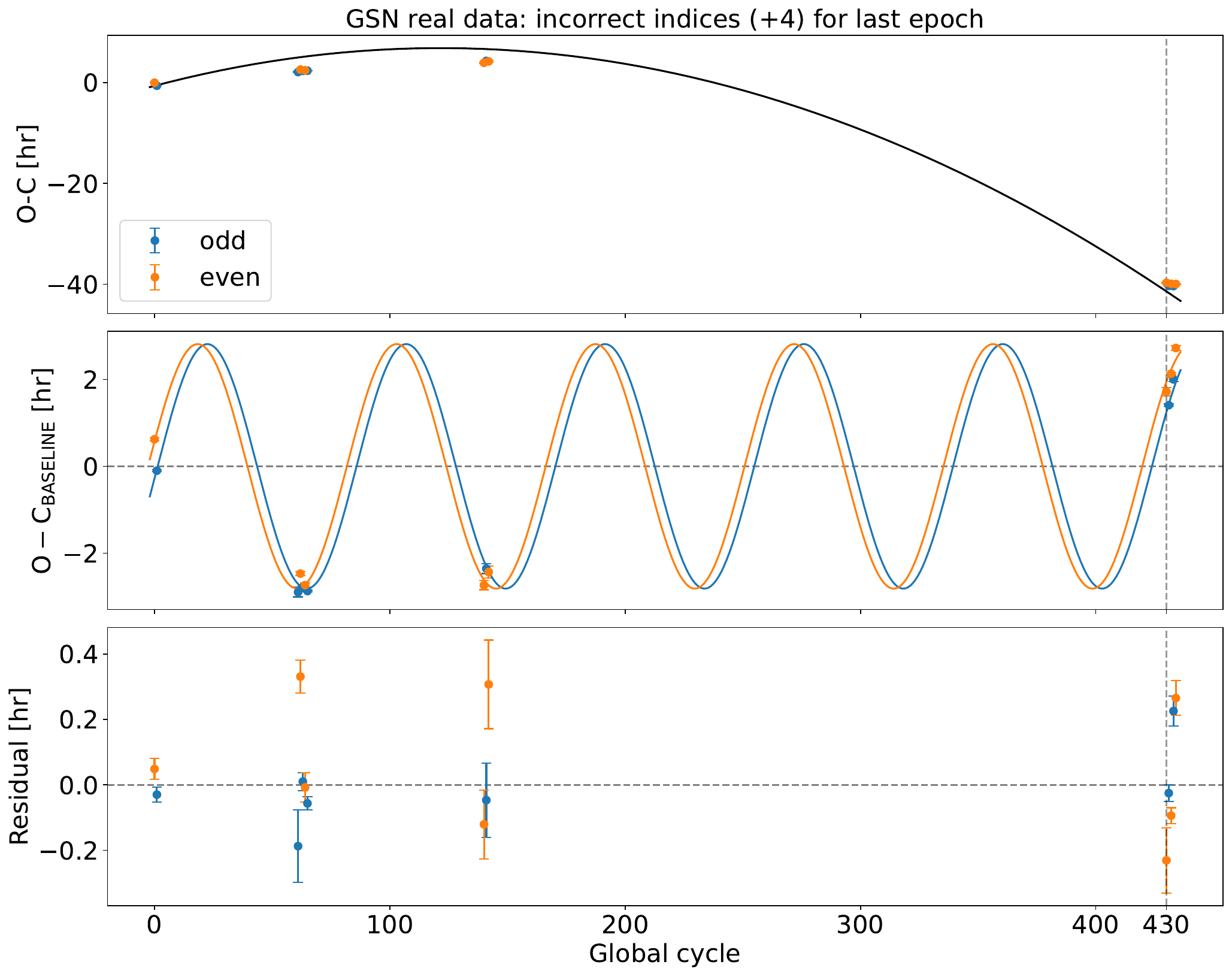}
    \includegraphics[width=0.45\linewidth]{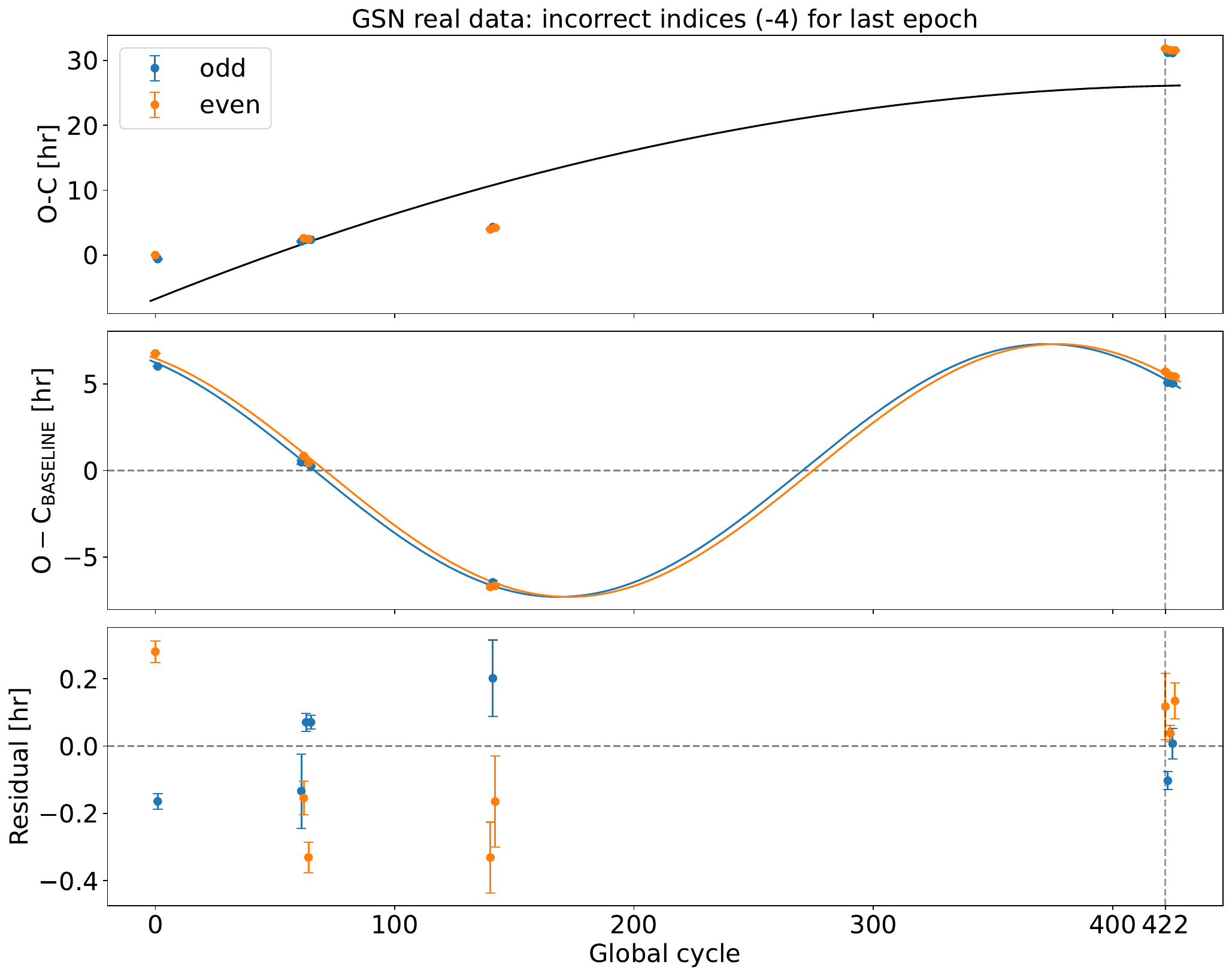}
    \caption{O-C analysis of GSN 069.  Upper panel: the cycle assignment favored by EMRI + disk model\cite{Zhou2025}, $N_{\rm FinalEpochStart}=426$.  Lower panels: two shifted cycle assignments, $N_{\rm FinalEpochStart}=430$ and $422$ (the one adopted in Ref.~\cite{Miniutti2025}).}
    \label{fig:real_data}
\end{figure}

\emph{Results.}
For $N_{\rm FinalEpochStart}=426$, the O-C diagram reveals an {\bf anti-phase} modulation between even and odd eruptions, with an amplitude about $0.3\ {\rm h}$.  This is the expected signature of apsidal precession in an eccentric EMRI crossing an equatorial disk.

For $N_{\rm FinalEpochStart}=430$ or $422$, the O-C diagram instead shows a {\bf large in-phase} modulation.  Its amplitude is roughly ten times larger than the anti-phase modulation amplitude.  The in-phase modulation could be interpreted as disk precession or another long-period modulation.  However, the mock tests above show that a large in-phase modulation could be a result of  an incorrect cycle assignment.

~\\

\emph{Which O-C is favored?}
There are two independent reasons to favor the $N_{\rm FinalEpochStart}=426$ solution.
\begin{enumerate}[leftmargin=*]
    \item The observed alternating long-short timing pattern $\Big($alternating $T_{\rm long}, T_{\rm short}$ $\&$ $T_{\rm long, short}(t)$ are time-dependent  $\&$ $T_{\rm long}(t)+T_{\rm short}(t)\approx$ const $\Big)$ is equivalent to an \emph{anti-phase} modulation in even and odd data.  O-C  that fail to recover this pattern are incorrect.

   In Fig.~\ref{fig:gsn_rec}, we plot the recurrence times $T_{\rm rec}(t)$ of GSN 069 eruptions. There are clearly two \emph{anti-phase} branches in $T_{\rm rec}(t)$. Each branch can be well fitted by a lin+sinusoidal function with period $\approx 76$ d.

   \begin{figure}[h]
    \centering
    \includegraphics[width=0.95\linewidth]{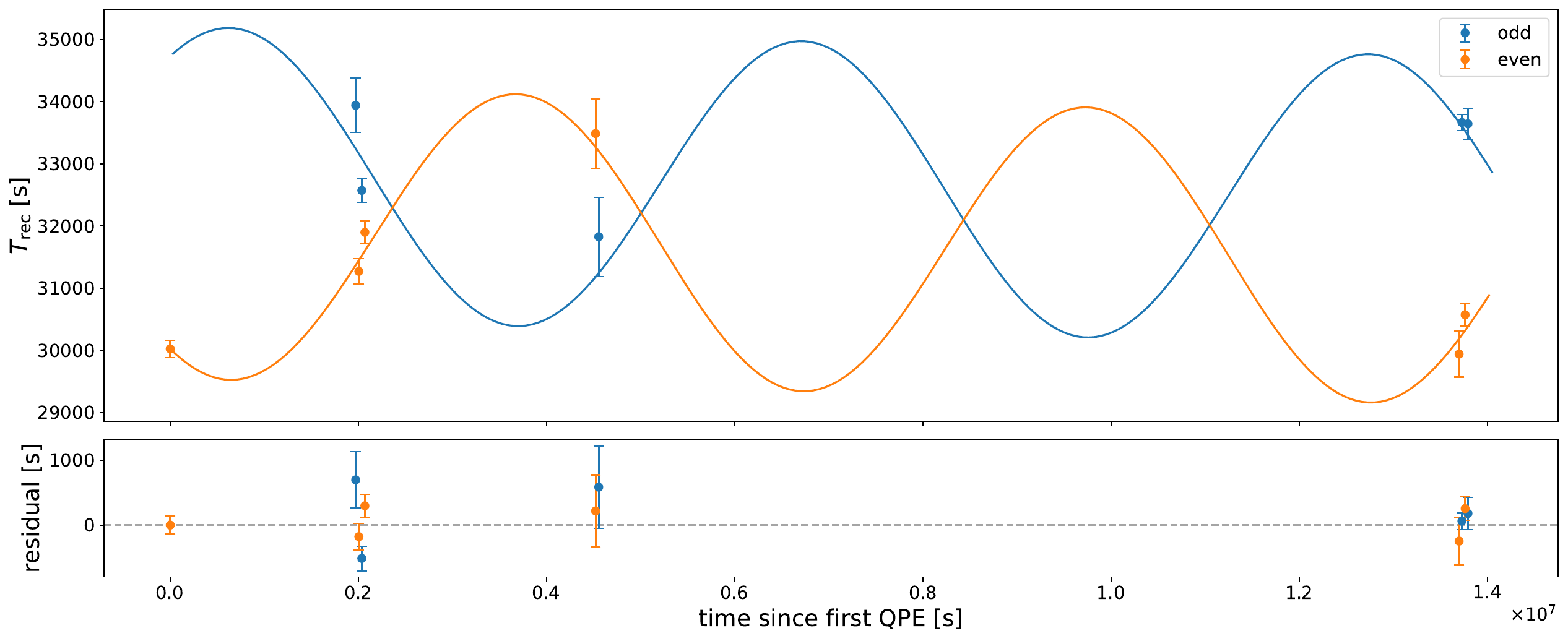}
    \caption{Recurrence times $T_{\rm rec}$ of GSN 069.  Blue and orange points show the odd and even $T_{\rm rec}$ data, respectively. 
    It is straightforward to see two \emph{anti-phase} branches in $T_{\rm rec}$, without using any O-C fit or cycle indexing. }
    \label{fig:gsn_rec}
\end{figure}

Without further information, there is uncertainty in the parity of the data in the last epoch due to data gaps. If the parity of the data in the last epoch is switched, the \emph{anti-phase} phase modulation in even and odd eruptions remains except with a longer period. 

\begin{figure}[h]
    \centering
    \includegraphics[width=0.95\linewidth]{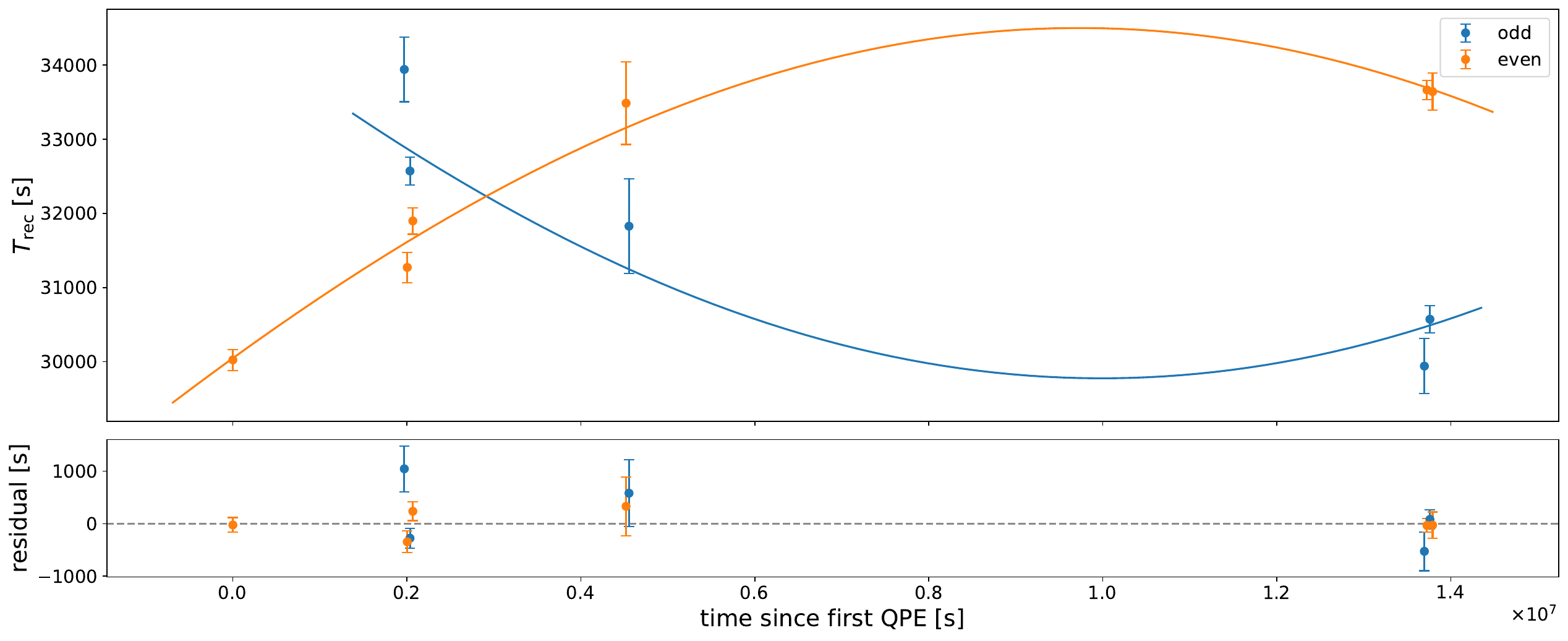}
    \caption{Lin+sinusoidal fit to the last-epoch-parity-switched data.}
\end{figure}

    \item When both apsidal precession and disk precession are included in the EMRI + disk model, we found the apsidal precession period $T_{\rm aps}=76^{+14}_{-34}\ {\rm d}$ ($2\sigma$), while the disk-precession period is much longer than the observation span.
\end{enumerate}

\emph{Conclusions.}

For GSN 069, both the well-known timing pattern (Fig.~\ref{fig:gsn_rec}) and correct O-C support an \emph{anti-phase} modulation rather than a large in-phase modulation ($\phi_{\rm even}-\phi_{\rm odd}=0$ is excluded at $>3\sigma$ C.L.).  In the EMRI + disk model, the observed anti-phase modulation is therefore evidence for apsidal precession, not for disk precession or another SMBH.

\newpage

\section*{VI. eRO-QPE2 real data from arXiv:2604.09788}

\phantomsection
\subsection*{VI(a). O-C toy-model fits}\label{subsec:ero2_oc}

% \emph{Purpose of this subsection.}
% The central issue for eRO-QPE2 in arXiv:2604.09788 is the cycle assignment of the last XMM epoch.  We compare $N_{\rm FinalEpochStart}=324$ and $323$ using several O-C toy models.  The goal is not to claim that a toy O-C model is the final physical model, but to check whether the data themselves really prefer the $323$ assignment.

% {\bf (a) O-C}

% Quad+mod: Following arXiv:2604.09788, we  fit the timing data with a quadratic polynomial + a sinusoidal modulation (quad + mod) for both 
% $N_{\rm FinalEpochStart}=324$ and 323.  The former reveals  $\dot T \approx -5\times10^{-5}$, and the latter reveals $\dot T \approx +3\times10^{-5}$. When comparing with the period decay trend in the archival data, the positive period derivative is excluded.

\emph{Quad + mod.}
Following Ref.~\cite{arcodia2026precessingclockrighttwice}, we first fit the timing data with a quadratic polynomial plus one sinusoidal modulation, denoted quad+mod.  For $N_{\rm FinalEpochStart}=324$, the fit gives a negative period derivative, $\dot T\approx -5\times10^{-5}$.  For $N_{\rm FinalEpochStart}=323$ (the one adopted in Ref.~\cite{arcodia2026precessingclockrighttwice}), the fit gives a positive period derivative, $\dot T\approx +3\times10^{-5}$.  The positive value is excluded when compared with the period decay trend in the archival data.

\begin{figure}[h]
    \centering
    \includegraphics[width=0.45\linewidth]{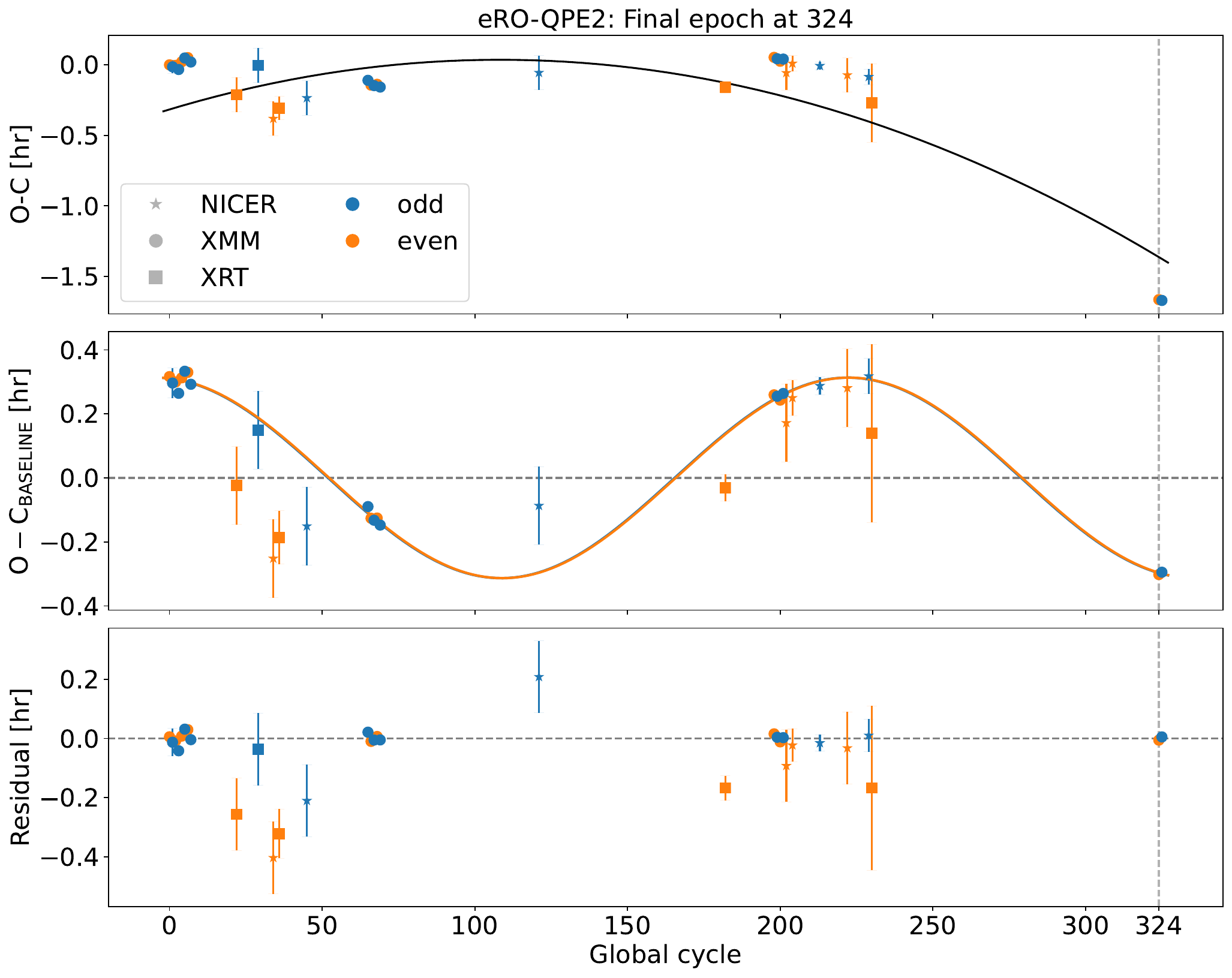}
    \includegraphics[width=0.45\linewidth]{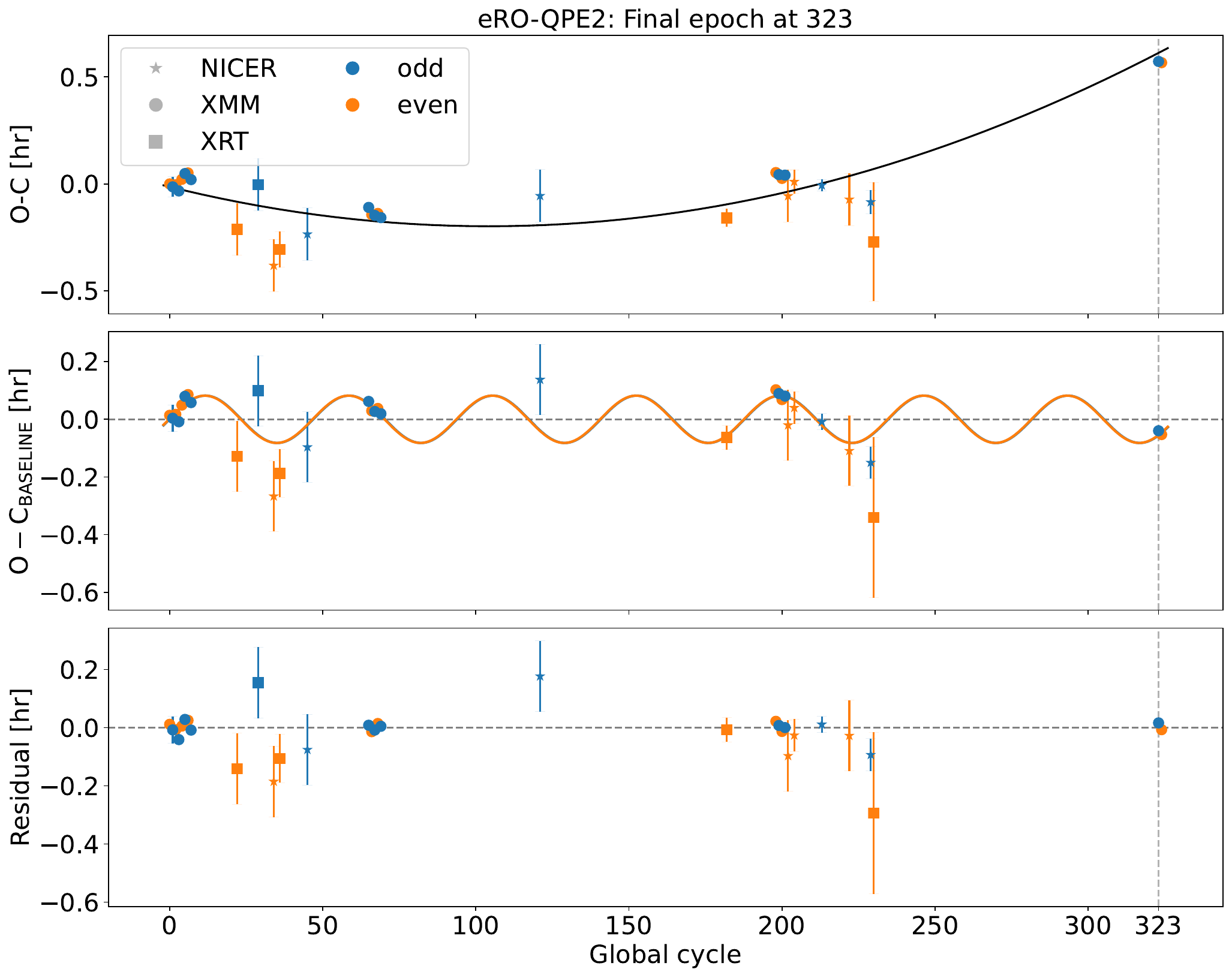}
    \caption{ Quad+mod fit to eRO-QPE2 timing data. Left: O-C with $N_{\rm FinalEpochStart}=324$, $\dot T = -5.49^{+0.32}_{-1.16}\times10^{-5} $ ($2\sigma$ C.I.). Right: O-C with $N_{\rm FinalEpochStart}=323$, $\dot T = +2.89^{+0.24}_{-0.32}\times10^{-5} $ ($2\sigma$ C.I.).}
    \label{fig:eRO2_bestfit_oc}
\end{figure}

% Lin+mod+mod: We  fit the timing data with a linear relation + two sinusoidal modulations for both 
% $N_{\rm FinalEpochStart}=324$ and 323. Again, no preference for 323 is found.

\emph{Lin + mod + mod.}
We also fit the data with a linear baseline plus two sinusoidal modulations, denoted lin+mod+mod.  This test is useful because a long-period sinusoid can partly mimic a quadratic trend over a finite observation span.  This model again shows no preference for $N_{\rm FinalEpochStart}=323$.

\begin{figure}[h]
    \centering
    \includegraphics[width=0.45\linewidth]{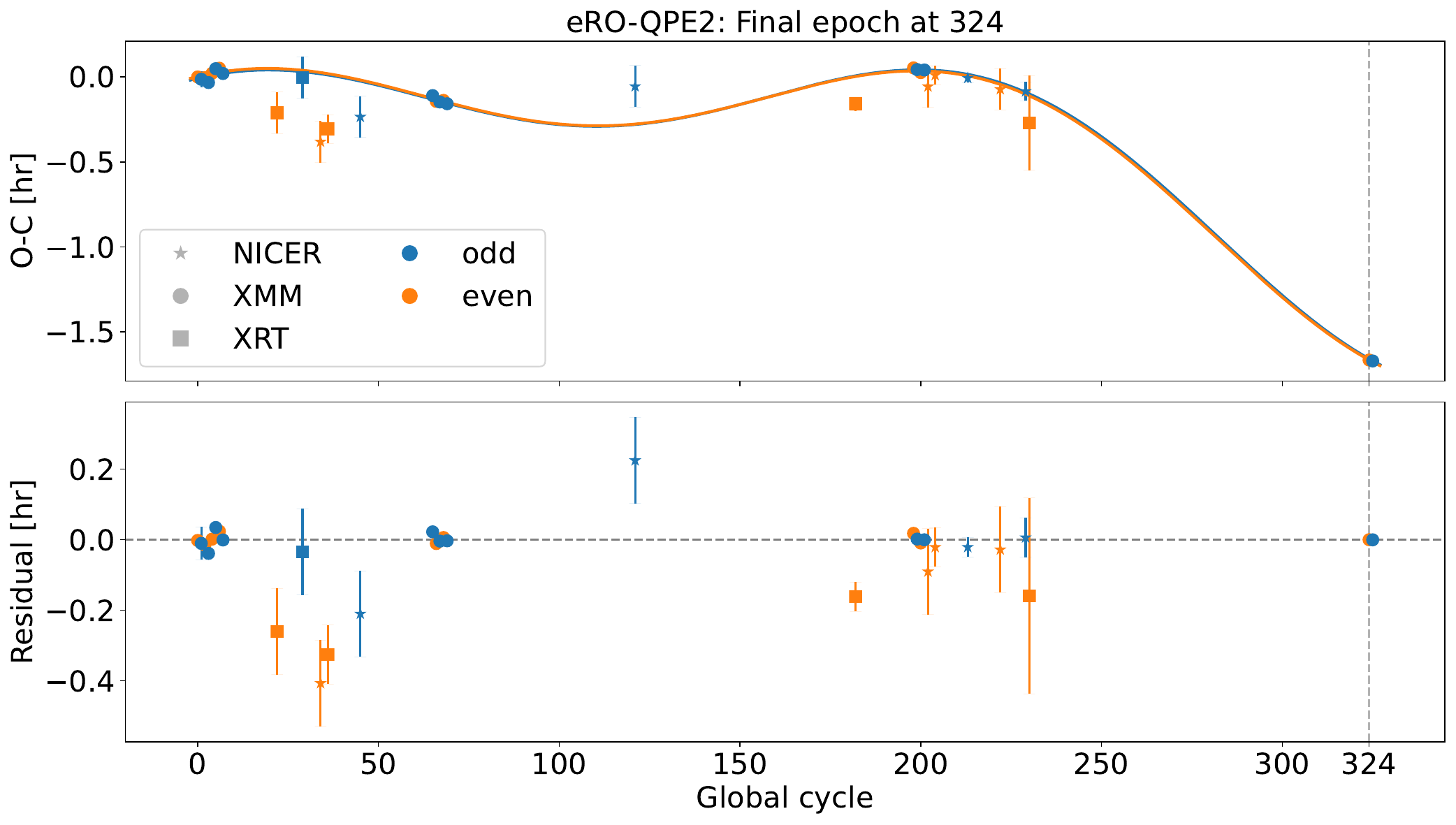}
    \includegraphics[width=0.45\linewidth]{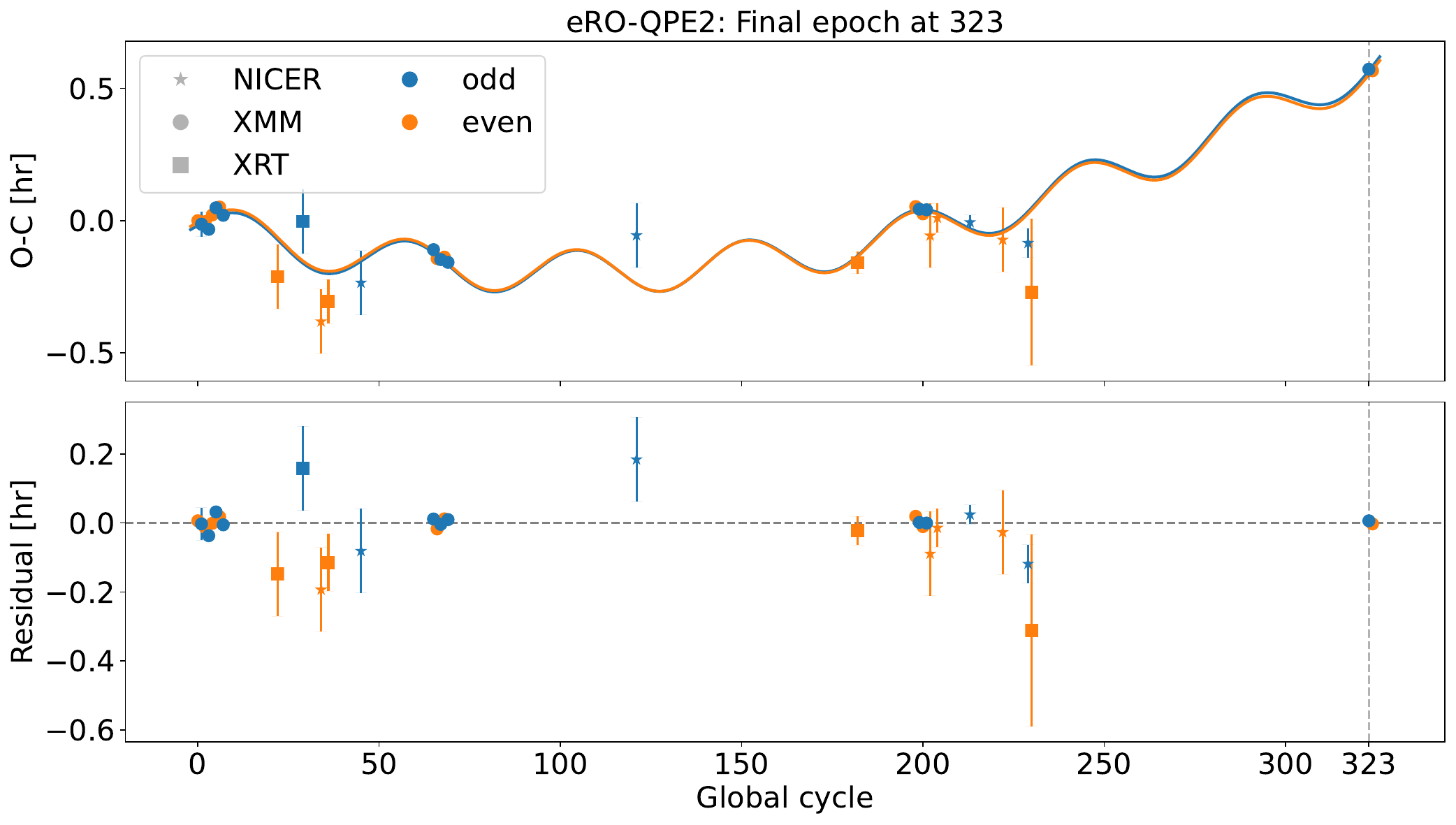}
    \caption{ Lin+mod+mod fit to eRO-QPE2 timing data. Left: O-C with $N_{\rm FinalEpochStart}=324$, $P_1=20.4^{+2.6}_{-2.7}\ {\rm d}, P_2=51.3^{+27.7}_{-18.3}\ {\rm d}$($2\sigma$ C.I.). Right: O-C with $N_{\rm FinalEpochStart}=323$, $P_1=4.4^{+0.1}_{-0.1}\ {\rm d}, P_2=118.1^{+31.4}_{-54.9}\ {\rm d}$($2\sigma$ C.I.).}
    \label{fig:eRO2_bestfit_2mod}
\end{figure}

% Quad+mod+mod: 
% the above two O-C analyses show that the data can be well fitted by either {\bf quad+mod} or {\bf lin+mod+mod} for both $N_{\rm FinalEpochStart}=324$ or 323. The reason is that the quadratic term 
% % \[ \frac{1}{2}\frac{\dot T}{T} t_i ^2 \]
% can be mimicked by a long-period sinusoid modulation. 
% \[A_2 \cos\left(\frac{2\pi t_i}{P_2}\right) = A_2 \left( 1 -\frac{1}{2}\frac{(2\pi t_i)^2}{P_2^2} + \mathcal{O}\left(\frac{(2\pi t_i)^4}{P_2^4}\right)\right)\]
% where $t_i(N_{\rm cyc}) = (N_{\rm cyc}-1)T$.  
% Though the timing data may be able to distinguish the quadratic term and the 
% long-period sinusoid modulation if the period is not much longer than the observation span.

\emph{Quad + mod + mod.}
The previous two fits show that the data can be fitted either by {\bf quad+mod} or by {\bf lin+mod+mod}.  This is expected because a long-period sinusoid can behave like a quadratic function over a limited time interval.

We fit the data with quad+mod+mod for both $N_{\rm FinalEpochStart} = 323 \ \& \ 324 $. Without imposing constraining prior information, we use a wide uniform prior, $\dot T\in\mathcal{U}[-5,5]\times10^{-4}$.  For $N_{\rm FinalEpochStart}=323$, we find  $\dot T=2.8^{+0.3}_{-0.4}\times10^{-5}$ ($2\sigma$) and $A_2< 300\ {\rm s}$ ($3\sigma$).  For $N_{\rm FinalEpochStart}=324$, we find $\dot T=-5.5^{+0.4}_{-1.1}\times10^{-5}$ ($2\sigma$)  and $A_2 < 500 \ {\rm s}$ ($3\sigma$).  Thus,  the data prefer a non-zero $\dot T$ rather than a  long-period-sinusoid mimic in the toy quad+mod+mod model.

% We fit the timing data with quad+mod+mod for both $N_{\rm FinalEpochStart}=324$ and 323.  Uniform priors are used, including $\dot T \in \mathcal{U}[-5,5]\times10^{-4}$.  For $N_{\rm FinalEpochStart}=323$, we find that $\dot{T}$ remains positive and is constrained to be $\dot{T}=2.8^{+0.3}_{-0.4}\times10^{-5}$ (2-$\sigma$), similar to the quad+mod fit. For $N_{\rm FinalEpochStart}=324$, $\dot{T}$ is, not surprisingly, negative with $\dot{T}=-5.5^{+0.4}_{-1.1}\times10^{-5}$ (2-$\sigma$), similar to the quad+mod fit.   This shows that the timing data favor a non-zero $\dot T$ instead of the mimic (at least for $P_2$ in the imposed prior range). 

% \begin{figure}[h]
%     \centering
%     \includegraphics[width=0.45\linewidth]{ero2_paper_324_OC_quadratic_sin1sin2_all_2row.pdf}
%     \includegraphics[width=0.45\linewidth]{ero2_paper_OC_quadratic_sin1sin2_all_2row.pdf}
%     \caption{ Quad+mod+mod fit to eRO-QPE2 timing data. Left: O-C with $N_{\rm FinalEpochStart}=324$, $P_1=21.9^{+3.5}_{-1.8}\ {\rm d}, P_2=98.3^{+50.4}_{-60.6}\ {\rm d}$(2-$\sigma$ C.L.). Right: O-C with $N_{\rm FinalEpochStart}=323$, $P_1=4.4^{+0.1}_{-0.1}\ {\rm d}, P_2=102.9^{+45.1}_{-60.6}\ {\rm d}$(2-$\sigma$ C.L.).}
%     \label{fig:eRO2_bestfit_quad2mod}
% \end{figure}

\emph{A negative-prior for $N_{\rm FinalEpochStart}=323$.}
 Data favor a positive $\dot T$ for $N_{\rm FinalEpochStart}=323$. If a negative prior is enforced, $\dot T<0$ with $\log_{10}(-\dot T)\in\mathcal{U}[-8,-4]$, the posterior is
artificially forced to the prior boundary with an apparent upper limit $-\dot T\lesssim10^{-6}$.  %This behavior is not evidence that the data prefer a tiny negative $\dot T$.  %It is a prior-boundary artifact caused by forcing the fit into a sign that the data do not prefer.

~\\

\emph{Conclusion from the O-C toy models.}
\begin{enumerate}
    \item O-C analyses show no preference for $N_{\rm FinalEpochStart}=323$.
    \item In the toy quad+mod+mod model, a negative posterior $-\dot T \lesssim 10^{-6}$ for $N_{\rm FinalEpochStart}=323$ appears to be a prior artifact. It arises only when a negative prior is enforced, even though the data themselves favor a positive $\dot T$. The posterior then peaks artificially at the prior boundary. 
    \item $N_{\rm FinalEpochStart}=323\ \& \ -\dot T \lesssim 10^{-6}$ is a strongly biased interpretation of the data. 
    %The combination $N_{\rm FinalEpochStart}=323$ and $-\dot T\lesssim10^{-6}$ is a biased interpretation produced by an imposed negative prior, not by the timing data alone.
\end{enumerate}

% Conclusions:
% \begin{enumerate}
%     \item O-C analyses show no preference for $N_{\rm FinalEpochStart}=323$.
%     \item In the toy quad+mod+mod model, a negative posterior $-\dot T \lesssim 10^{-6}$ for $N_{\rm FinalEpochStart}=323$ appears to be a prior artifact. It arises only when a negative prior is enforced, even though the data themselves favor a positive $\dot T$. The posterior then peaks artificially at the prior boundary.
%     \item $N_{\rm FinalEpochStart}=323\ \& \ -\dot T \lesssim 10^{-6}$ is a strongly biased interpretation of the data.
% \end{enumerate}

\newpage

% {\bf (b) EMRI+disk}

\phantomsection
\subsection*{VI(b). EMRI + disk}\label{subsec:ero2_emri}

\begin{figure}[h]
    \centering
    \includegraphics[width=\linewidth]{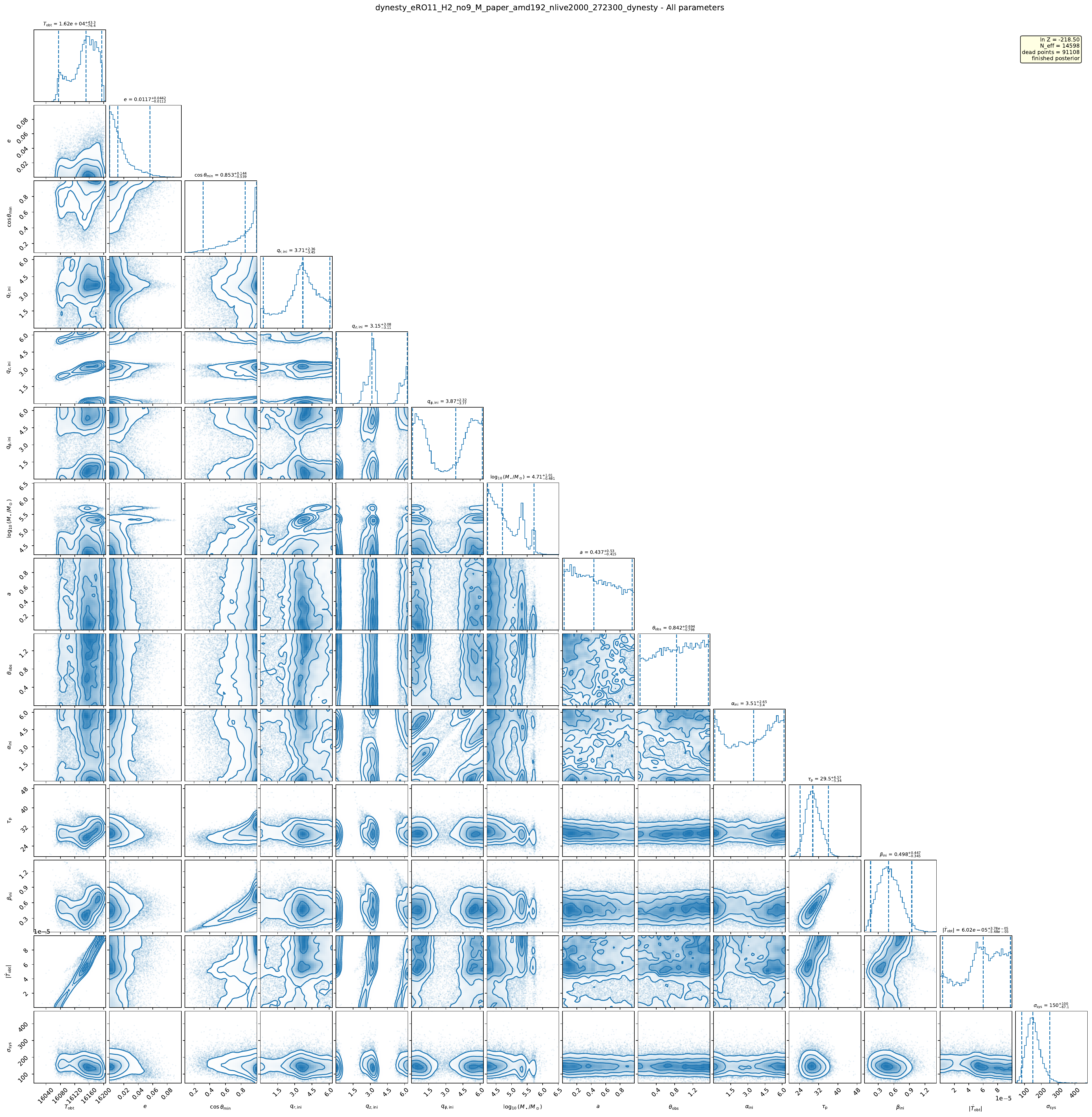}
    \caption{Corner plot of posterior samples for the EMRI+disk model fit of the eRO-QPE2 data.}
    \label{fig:eRO2_corner}
\end{figure}

We fit the eRO2 timing data in Ref.~\cite{arcodia2026precessingclockrighttwice} in the framework of EMRI+disk model. The posterior constraints are shown in Fig.~\ref{fig:eRO2_corner},
where the EMRI is near-circular with $e$ peaking at 0, the disk is precessing with $\tau_{\rm p} = 29.5^{+6.57}_{-5.34} $ d ($2\sigma$),
and the EMRI orbital period decay rate $\dot T_{\rm obt}= -6.0^{+5.8}_{-4.7}\times 10^{-5}$ ($2\sigma$). Note that $\dot T_{\rm obt}$ constraint in EMRI+disk is different from the $\dot T$ constraints in O-C toy models (quad+mod, quad+mod+mod). They are different model parameters in different models.

~\\

\begin{figure}[h]
    \centering
    \includegraphics[width=0.45\linewidth]{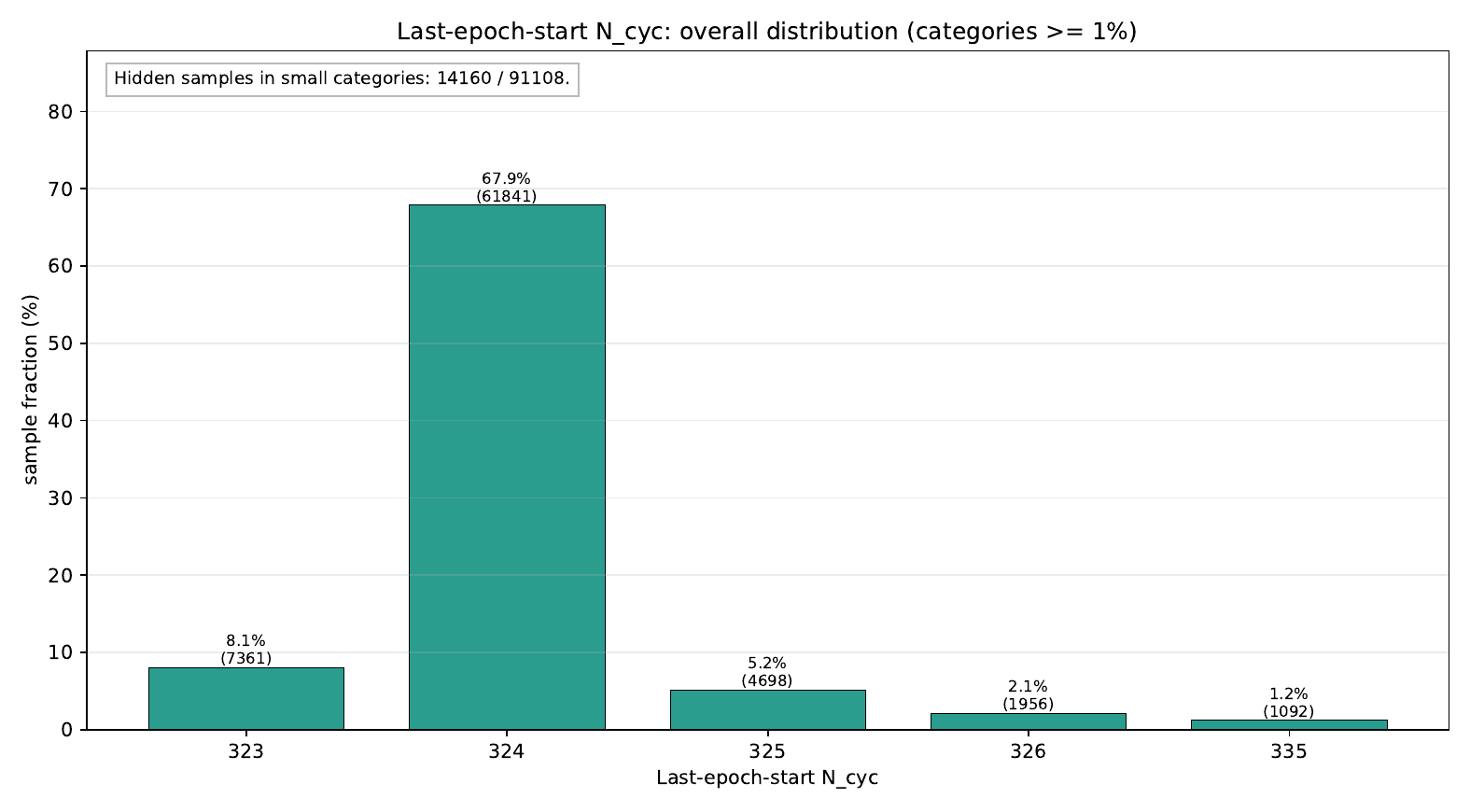}
    \includegraphics[width=0.45\linewidth]{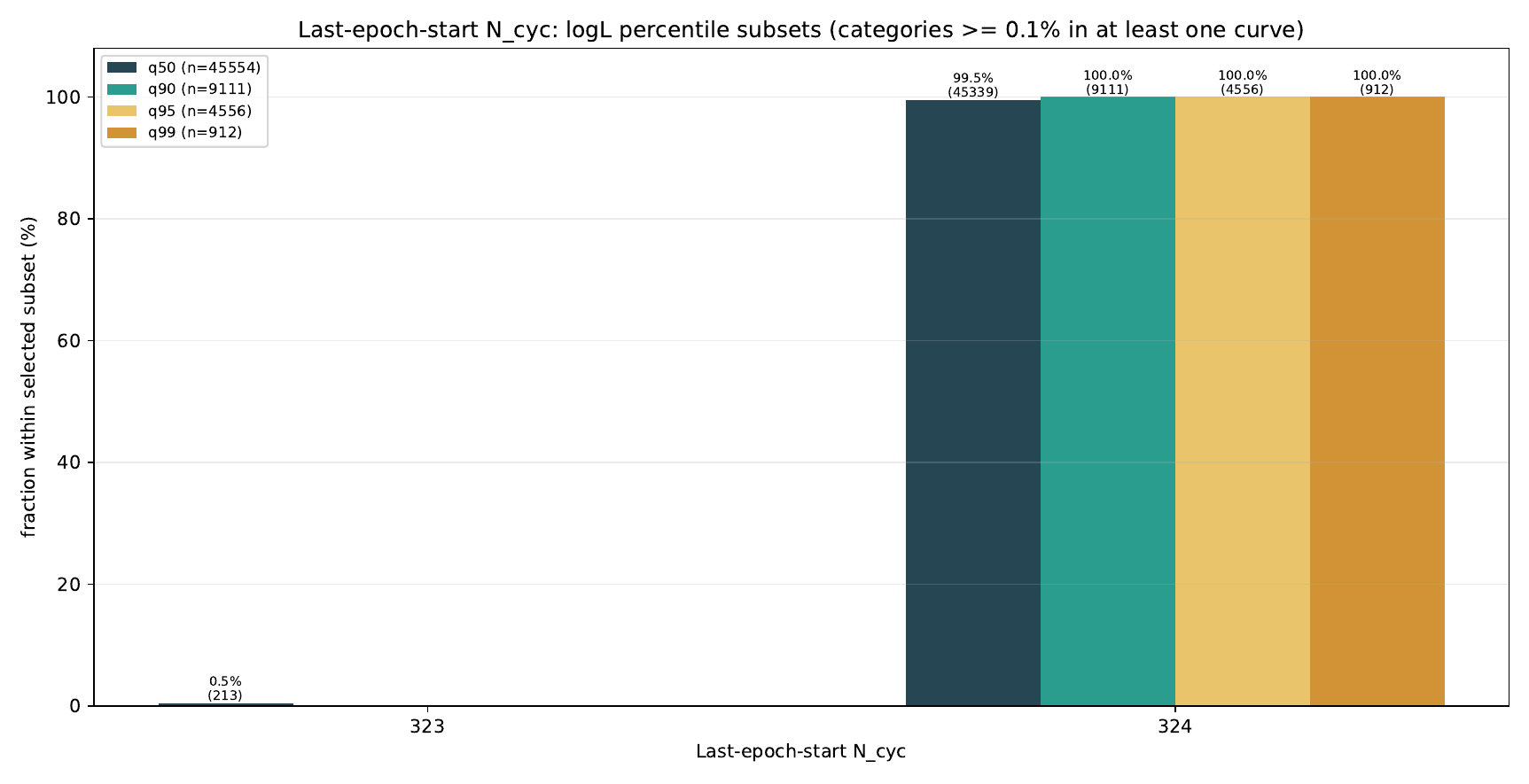}
    \caption{Posterior distribution of the starting $N_{\rm cyc}$ of the last epoch $N_{\rm FinalEpochStart}$ among
    all valid dynesty samples (top 99\% by posterior weight). $N_{\rm FinalEpochStart}=324$ is
    strongly preferred over $323$. }
    \label{fig:ncyc_overall}
\end{figure}

\emph{$N_{\rm cyc}$ inference.}
In the Bayesian inference of EMRI+disk model parameters, the QPE cycle numbers $N_{\rm cyc}$ are inferred from data instead of being prefixed. Figure~\ref{fig:ncyc_overall} shows the posterior distribution of the starting cycle number of the last epoch, $N_{\rm FinalEpochStart}$. The result is not single-valued, because there is some uncertainty in {\bf $N_{\rm cyc}$ } due to data gaps.  However, $N_{\rm FinalEpochStart}=324$ is clearly favored over other possibilities.  This is a Bayesian inference result from the data, instead of a prior imposed.

% In the Bayesian inference of EMRI+disk model parameters, the QPE cycle numbers $N_{\rm cyc}$ are inferred from data instead of being prefixed. In Fig.~\ref{fig:ncyc_overall}, we show the constraint of $N_{\rm FinalEpochStart}$. It is of no surprise to see $N_{\rm FinalEpochStart}$ is not single-valued, and  $N_{\rm FinalEpochStart}=324$ is clearly favored over other possibilities. Note that this is a result of Bayesian inference, instead of a prior imposed. 

\begin{figure}[h]
    \centering
    \includegraphics[width=0.45\linewidth]{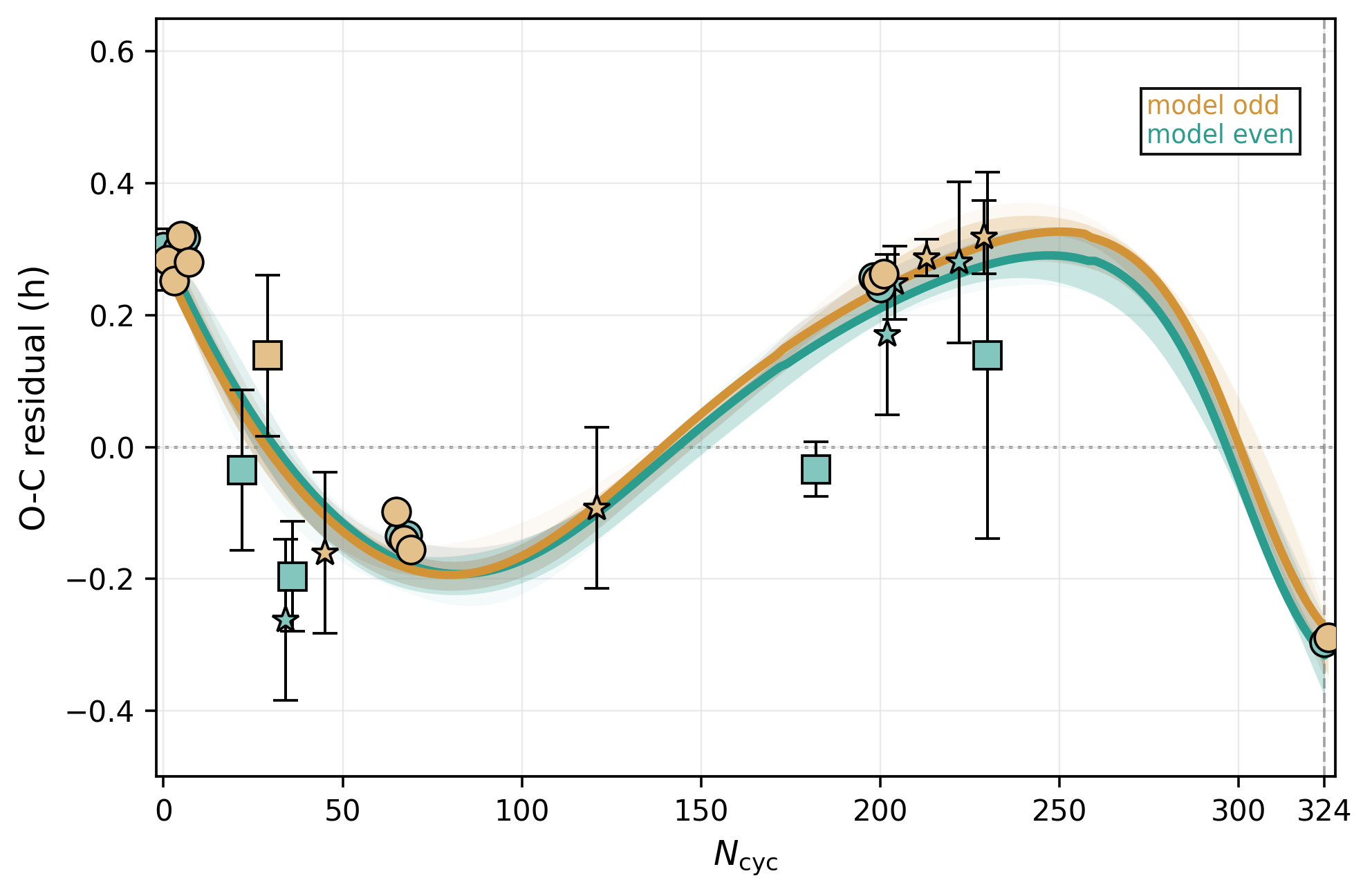}
    \caption{Applying O-C to mock data generated from posterior samples of the EMRI+disk model fit in Fig.~\ref{fig:eRO2_corner}.}
    \label{fig:eRO2_mock_oc}
\end{figure}

\emph{O-C consistency check.}
With $N_{\rm FinalEpochStart}=324$, the O-C fit shows a quadratic baseline plus an in-phase modulation with period $P=22^{+4}_{-2}\ {\rm d}$ ($2\sigma$ C.I.) (left panel of Fig.~\ref{fig:eRO2_bestfit_oc}). This is consistent with the disk precession period inferred in the EMRI + disk model.  To further check whether the fit is not a numerical artifact, we generate mock timing data with top $10\%$ high-likelihood posterior samples and apply the same O-C  to the mock data.  The mock data reproduce the same qualitative behavior, as shown in Fig.~\ref{fig:eRO2_mock_oc}. No numerical artifacts is found.

% With $N_{\rm FinalEpochStart}=324$, O-C shows that the data can be fitted by quadratic + an in-phase sinusoidal modulation in even and odd data with a period $P= 22^{+4}_{-2} $ d ($2\sigma$ C.I.) (left panel of Fig.~\ref{fig:eRO2_bestfit_oc}). This modulation is well consistent with the disk precession as inferred in the framework of EMRI+disk.  As a double check of the EMRI+disk model fitting result, we generate mock data with  top $10\%$ best fit parameter samples and apply the same O-C to the mock data (Fig.~\ref{fig:eRO2_mock_oc}). No numerical artifact is found.

\begin{figure}[h]
    \centering
    \includegraphics[width=0.45\linewidth]{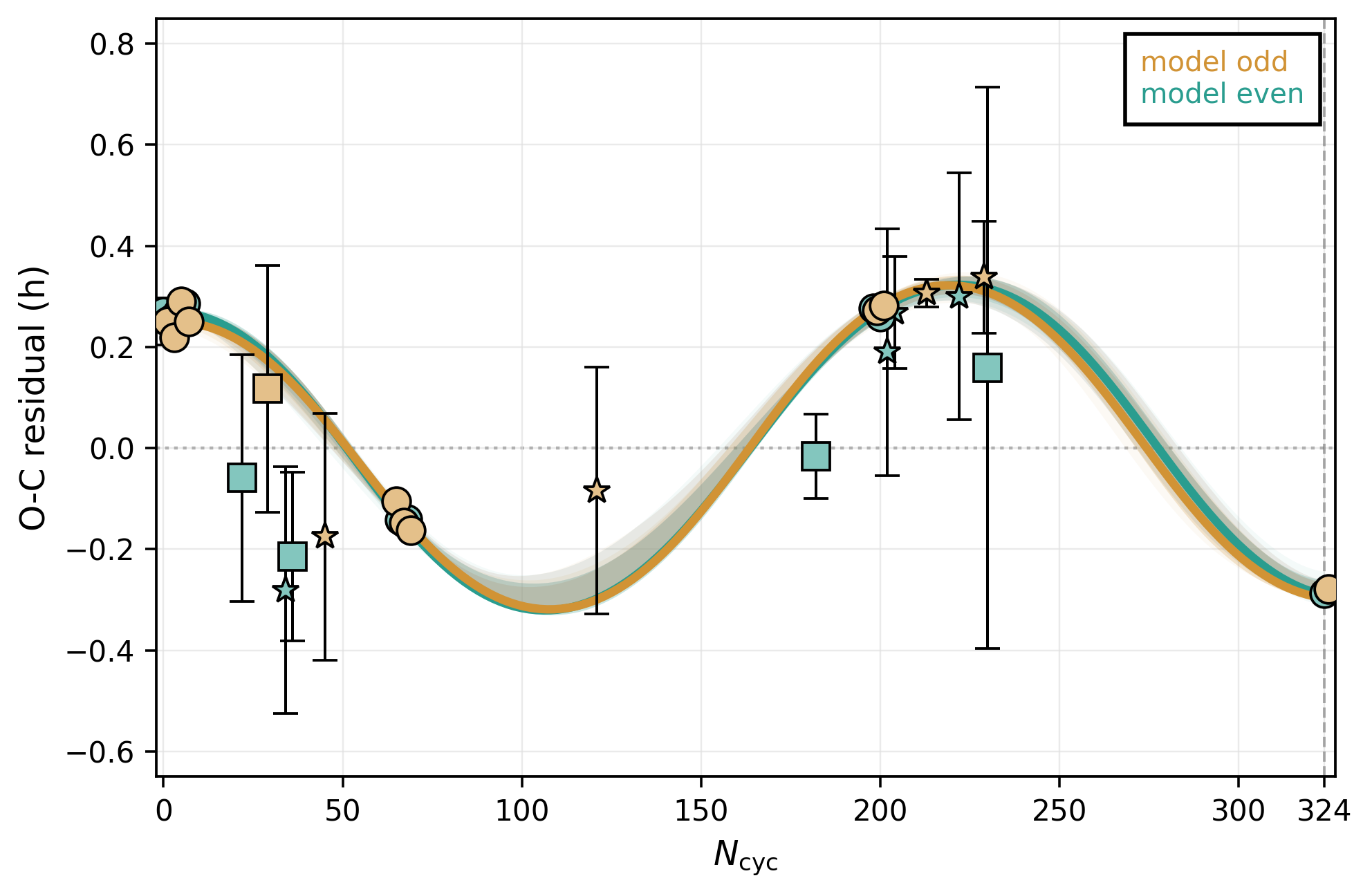}
    \caption{Applying O-C to mock data generated from posterior samples of the EMRI+disk model fit with more conservative error bars assigned to NICER and XRT timing points.}
    \label{fig:eRO2_mock_oc2}
\end{figure}

In  Fig.~\ref{fig:eRO2_mock_oc},  a few timing points deviate from the EMRI+disk solution indicating a potential issue in the fit. 
In practice, we find the goodness of fit depends strongly on the error bars {\bf assigned} to the few timing points where sometimes only a single data point is available to constrain a given flare timing. The fit improves considerably when larger error bars are adopted for those points (see Fig.~\ref{fig:eRO2_mock_oc2}).
With more conservative error bars assigned to the timing points of NICER and XRT, the constraints of model parameters remain largely unaffected, except  both the median value and uncertainty of $\sigma_{\rm sys}$ are reduced.

% \begin{figure}[h]
%     \centering
%     \includegraphics[width=\linewidth]{corner_double.jpg}
%     \caption{Same to Fig.~\ref{fig:eRO2_corner} except two times larger error bars are adopted for NICER and XRT data points.}
%     \label{fig:eRO2_corner_doulble}
% \end{figure}

~\\

\emph{Conclusion from the EMRI + disk fit.}
\begin{enumerate}
    \item The eRO2 data are well modeled by a near-circular EMRI + a precessing disk with an orbital period decay rate $\dot T_{\rm obt}= -6.0^{+5.8}_{-4.7}\times 10^{-5}$ (2-$\sigma$). 
    \item The Bayesian inference accounts for uncertainties in $N_{\rm cyc}$ self-consistently. The posterior shows support for a number of possibilities of $N_{\rm FinalEpochStart}$, as it should be (see Figure~\ref{fig:ncyc_overall}).
    \item No numerical artifacts is found in the O-C consistency check.
\end{enumerate}

\newpage 

% {\bf (c) Enforcing $N_{\rm FinalEpochStart}=323\ \& \ \dot T_{\rm obt} =0 $ on EMRI+disk}

\subsection*{VI(c). Enforcing $N_{\rm FinalEpochStart}=323$ and $\dot T_{\rm obt}=0$ on EMRI+disk}

O-C analyses in \hyperref[subsec:ero2_oc]{(a)}  show $N_{\rm FinalEpochStart}=323\ \& \ -\dot T \lesssim 10^{-6} $ is a strongly biased  interpretation of the data. Bayesian inference in \hyperref[subsec:ero2_emri]{(b)} shows  that the eRO2 timing data are well modeled by EMRI+disk. What if we fit the data (XMM 1-4) with the EMRI+disk model with an extra constraint $N_{\rm FinalEpochStart}=323\ \& \ \dot T_{\rm obt} =0 $ imposed ? In this setup, 
the nominal EMRI+disk solution obtained was found  to be  merely numerical artifacts (Figure~C. of Ref.~\cite{arcodia2026precessingclockrighttwice}).

As a comparison, we perform Bayesian inferences of EMRI+disk with two different hypotheses: (1)  free  $N_{\rm cyc}$ and $\dot T_{\rm obt}$, (2) fixed $N_{\rm FinalEpochStart}=323\ \& \ \dot T_{\rm obt} =0 $. The log Bayes factor $\log \mathcal{B}_{\rm fixed}^{\rm free} = 11.5$ obtained excludes hypothesis (2) decisively. 

~\\

\emph{Results.}
% Summary of (b) and (c):
\begin{enumerate}
    \item (EMRI+disk) + data + (free  $N_{\rm cyc}$ and $\dot T_{\rm obt}$) $\Rightarrow$ no numerical artifacts
    \item (EMRI+disk) + data + (fixed $N_{\rm FinalEpochStart}=323\ \& \ \dot T_{\rm obt} =0 $) $\Rightarrow$ numerical artifacts
    \item  $\log \mathcal{B}_{\rm fixed}^{\rm free} = 11.5$.
\end{enumerate}

% \begin{figure}[h]
%     \centering
%     \includegraphics[width=0.7\linewidth]{new_plots/EMRIfit_oddeven_artifact.pdf}
%     \caption{Figure~C.1 of arXiv:2604.09788 (reproduced for reference).
%     The O-C diagram based on the best-fit orbit of arXiv:2604.09788 shows a small
%     {\bf anti-phase} oscillation between even and odd timings.}
%     \label{fig:paper_c1}
% \end{figure}

%Fig. Applying O-C to mock data for  fixing 323 and $\dot T_{\rm obt}=0$ (Figure~C.1 of arXiv:2604.09788)

\emph{Conclusions.}
\begin{enumerate}
    \item EMRI+disk is compatible with the eRO2 timing data, but is NOT compatible with the hypothesis $N_{\rm FinalEpochStart}=323\ \& \ \dot T_{\rm obt} =0$.
    Enforcing this constraint on EMRI+disk only leads to numerical artifacts.
    \item \emph{ Constrain a model with data.}  Do not constrain a model with data + anything beyond data, including extra constraints extracted from data with O-C models.  If the extra constraints are correct, double copies of information incorrectly suppress error bars of model parameters. If the extra constraints are biased, the inference is also biased.
\end{enumerate}

~\\

{\bf A  eRO2 summary:}
\begin{itemize}
    \item (a) O-C show $N_{\rm FinalEpochStart}=323\ \& \ -\dot T \lesssim 10^{-6}$ is a strongly biased interpretation of the data.
    \item (b) The data are well modeled by a near-circular EMRI+ a precessing disk with an orbital period decay rate $\dot T_{\rm obt}= -6.0^{+5.8}_{-4.7}\times 10^{-5}$ (2-$\sigma$).
    \item (c) EMRI+disk is compatible with the data, but not the hypothesis $N_{\rm FinalEpochStart}=323\ \& \ \dot T_{\rm obt} =0$.
\end{itemize}

\newpage

{\bf Summary}
\begin{enumerate}
    \item O-C analysis is highly sensitive to the cycle number  $N_{\rm cyc}$. A small mismatch leads to large false signals, and \emph{a universal signature of these false signals is a large in-phase sinusoidal modulation in even and odd data. } Therefore, uncertainties in $N_{\rm cyc}$ must be properly quantified for making valid claims based on O-C. 
    %, which has been mistaken as an EMRI orbiting around another SMBH in the literature. As a result, the prevalence of incorrect O-C leads to the prevalence of EMRIs in SMBH binaries.
    \item The GSN 069 data are well modeled by a low-eccentricity EMRI + an equatorial disk ($e\approx 0.04$), where the anti-phase modulation is driven by apsidal precession ($T_{\rm aps}\approx 76$ d) and there is no evidence for an in-phase modulation.
    Incorrect O-C  analyses with a small mismatch in $N_{\rm cyc}$ lead to large false alarms.
    \item The eRO-QPE2 data are well modeled by a near-circular EMRI + a precessing disk ($e \approx 0 , \tau_{\rm p}\approx 25$ d, and loosely bounded $\dot T_{\rm obt}$), where the anti-phase modulation caused by apsidal precession is small and the in-phase modulation caused by disk precession is dominant.
    Incorrect O-C  with a small mismatch in $N_{\rm cyc}$ lead to large false alarms.
\end{enumerate}

\bibliographystyle{apsrev4-1}
\bibliography{refs}

\end{document}